\documentclass{ws-ijtaf}

\usepackage{amsmath,amsfonts,amssymb, mathpazo}
\usepackage{latexsym,epsfig,graphics,mathrsfs}
\usepackage{dsfont,wrapfig,bbm, fixmath, bm,multirow}
\usepackage{comment}
\usepackage{placeins}

\usepackage{enumerate}
\usepackage{hyperref}
\usepackage{setspace}
\usepackage{subfigure}

\usepackage{algorithm}
\usepackage{algpseudocode}  
\usepackage{amsmath,amsfonts,amssymb,color,tikz}
\usepackage{booktabs}
\usepackage{threeparttable}  % Add this to your preamble
 \newcommand{\beginsupplement}{
    \setcounter{section}{0}
    \renewcommand{\thesection}{S\arabic{section}}
    \setcounter{equation}{0}
    \renewcommand{\theequation}{S\arabic{equation}}
    \setcounter{table}{0}
    \renewcommand{\thetable}{S\arabic{table}}
    \setcounter{figure}{0}
    \renewcommand{\thefigure}{S\arabic{figure}}
    \newcounter{SIfig}
    \renewcommand{\theSIfig}{S\arabic{SIfig}}}

\usepackage{mathrsfs}      
\usepackage{url}
\usepackage{multirow}
\usepackage[normalem]{ulem}
\useunder{\uline}{\ul}{}
  
\begin{document}

\markboth{Jiwon Jung, Kiseop Lee}{Attention-Based Reading, Highlighting, and Forecasting of the Limit Order Book}

\catchline{}{}{}{}{}
%%%%%%%%%%%%%%%%%%%%%%%%%%%%%%%%%%%%%%%%%%%%%%%%%%%%%%%%%%%%%%%%%%%%

\title{Attention-Based Reading, Highlighting, and Forecasting\\ of the Limit Order Book}

\author{Jiwon Jung\footnote{Corresponding author}}  
\address{Department of Statistics, Purdue University, West Lafayette, IN 47907, USA \\
\email{jung320@purdue.edu} 
}

\author{Kiseop Lee}  
\address{Department of Statistics, Purdue University, West Lafayette, IN 47907, USA \\
\email{kiseop@purdue.edu} 
}
    
% \author{SECOND AUTHOR}

% \address{Group, Laboratory, Address\\
% City, State ZIP/Zone, Country\\
% author\_id@domain\_name }

\maketitle

% \begin{history}
% \received{(Day Month Year)}
% \revised{(Day Month Year)}
% \end{history}
\begin{abstract}
Managing high-frequency data in a limit order book (LOB) is complex, often beyond the capabilities of conventional time-series models. Accurate prediction of the multi-level LOB, not just the mid-price, is crucial for understanding market dynamics but is difficult due to the interdependencies among attributes like order types, features, and levels. This study introduces advanced sequence-to-sequence models to forecast the entire multi-level LOB, including prices and volumes. Our key contribution is a compound multivariate embedding method that captures spatiotemporal relationships. Empirical results show that our method outperforms others, achieving the lowest forecasting error while maintaining LOB structure.
\end{abstract}


\section{Introduction}

A limit order book (LOB) is a digital record of all outstanding buy and sell orders for a particular financial instrument, such as a stock or currency. It contains information about the price and quantity of each order, as well as the time at which the order was placed. The LOB continuously updates as new orders are submitted, executed, or canceled, providing an up-to-the-moment snapshot of market supply and demand.  
This evolution of orders highlights the distribution of orders across various price levels, providing information on market depth and potential price movements.  
 
However, predictive modeling of LOBs presents considerable challenges due to the high frequency and complexity of the data. The dynamic nature of the LOB, with orders being added, modified, or removed at millisecond intervals, requires rapid processing and analysis of vast amounts of information. Additionally, LOB data has a complex multivariate time-series structure with compound attributes, where levels, types, and features are interrelated and simultaneously associated with each value. For instance, the ordinal structure in multi-level prices highlights the interconnectedness of these attributes. Therefore, accurately capturing the influence of each attribute is essential for effective predictive modeling of the LOB.

Another challenge in time series forecasting, including LOB prediction, is the non-stationarity of the data. Non-stationarity refers to changes in the statistical properties of a time series over time, such as mean, variance, or autocorrelation, which can significantly impact forecasting accuracy. Shifts in trends and structural changes over time can result in distributional differences between the training and test periods, undermining model performance. Techniques such as adaptive attention mechanisms \cite{liu2022non} and trainable normalization methods \cite{kim2021reversible} have been proposed to address these issues, but still struggle with abrupt changes in high-frequency data and require careful tuning to handle long-term dependencies.

\begin{figure}[pb]
\centerline{\includegraphics[width=\textwidth]{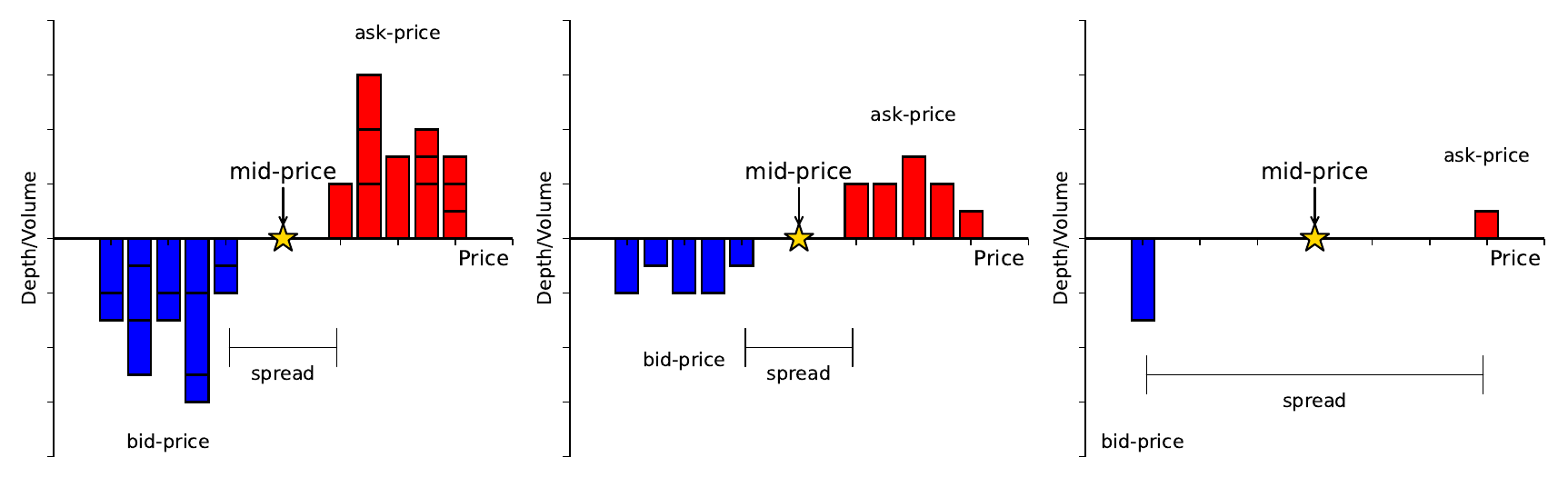}}
        % \centering
        % \includegraphics[width=\textwidth]{lob/three_lob.pdf}
        \caption{Three different LOB snapshots with the same mid-price: The three limit order book snapshots share the same mid-price, with the first and second snapshots showing varying levels of depth available, while the third snapshot features a larger bid-ask spread.} 
        \label{graphical_rep}
\end{figure}

Given the complexities of predicting the full dynamics of an LOB, many researchers have focused on forecasting the mid-price, defined as the average of the best bid and ask prices. Various machine learning models have been applied to this task, including regression-based approaches \citep{ntakaris2018benchmark}, support vector machines \citep{kercheval2015modelling}, and deep neural networks \citep{liu2024vit, tsantekidis2020using, zhang2019deeplob} to predict future mid-price movements. 
The mid-price is a general market trend indicator. However, Figure~\ref{graphical_rep} shows that the LOB can vary across features like order depths at different price levels and bid-ask spreads (BAS), even if the mid-price stays the same.
Thus, relying solely on mid-price analysis may be insufficient when a deeper understanding of order book depth is necessary. This limitation is also noted in \citet{sirignano2019deep}, where the authors emphasize the importance of modeling the joint distribution of bid and ask prices rather than assuming a constant bid-ask spread around the mid-price. Their approach expands mid-price classification to joint price distribution prediction, though bid and ask sizes are only used as inputs and not predicted jointly. A comprehensive view of the LOB, including bid and ask prices and their respective depths, is essential for risk management applications, such as in optimal execution, where fully capturing the LOB structure is crucial.

Attention-based models like Transformer \citep{vaswaniAttentionAllYoua} and BERT \citep{devlinBERTPretrainingDeep2019} excel at sequence-to-sequence prediction by assigning varying weights to different parts of the data, enabling them to focus on the most relevant information for more accurate predictions. These models have proven highly effective in tasks such as natural language processing \citep{yang2021context, galassi2020attention, cui2019fine}, speech recognition \citep{zeyer2018improved, chorowski2015attention, bahdanau2016end}, and time series forecasting \citep{shih2019temporal, fan2019multi,du2020multivariate}. Despite their success, applications of attention-based models in limit order book forecasting are still relatively limited. Recently, however, studies have highlighted their potential in this area: \citet{arroyo2024deep} used survival analysis to forecast order fill probabilities, \citet{liu2024vit} applied vision transformers to classify mid-price movements, and \citet{wu2024mwdn} employed wavelet decomposition for short-term price movement prediction in the Chinese market. Nevertheless, none of these studies focused on predicting the full price and volume structure within the limit order book.

In this study, we present an attention-based approach utilizing the Spacetimeformer architecture \citep{grigsbyLongRangeTransformersDynamic2022} to address the unique challenges of limit order book (LOB) forecasting. Our method builds on previous research by introducing a customized embedding technique that effectively captures spatiotemporal features while maintaining the LOB’s structural integrity, resulting in a more detailed and accurate depiction of market dynamics. By integrating both price and volume information, this approach significantly enhances predictive accuracy compared to traditional models, showcasing the strong potential of attention-based architectures to advance forecasting in high-frequency trading contexts. These advancements highlight the critical role of attention mechanisms in identifying complex patterns within LOB data, paving the way for more sophisticated and reliable market modeling.

Our contributions are three-fold: (1) we extend the capability of attention-based models beyond mid-price prediction, enabling them to capture dynamic patterns within the entire LOB structure; (2) we implement modified spatiotemporal embeddings that effectively integrate both spatial and temporal features, offering a model generalizable to multivariate time series with compound attributes; and (3) we improve forecasting accuracy by addressing non-stationarity in high-frequency financial data using a percent-change and min-max transformation method that does not add complexity to the model.

This paper is organized as follows: Section 2 offers an overview of the dataset and experimental setup. Section 3 details the methods used to address the challenges of LOB forecasting. Section 4 presents the experimental results, comparing our model's performance with various existing techniques. Section 5 discusses the implications of our approach, and Section 6 concludes the paper with suggestions for future research directions.

% \section{Literature review}

\section{Data}  
 
The LOB is a real-time record of all buy and sell orders for a specific asset in financial markets. Organized by price levels, it displays the best bid (highest buy price) and best ask (lowest sell price) at any given moment, along with details like order volume and timestamps. In this section, we explore the high-frequency characteristics and spatiotemporal structure of the LOB data and outline our approach for formulating and transforming these inputs to support model development.

\subsection{High-frequency dynamics and spatiotemporal correlations}

\begin{figure} 
    \centering
    \includegraphics[width=\textwidth]{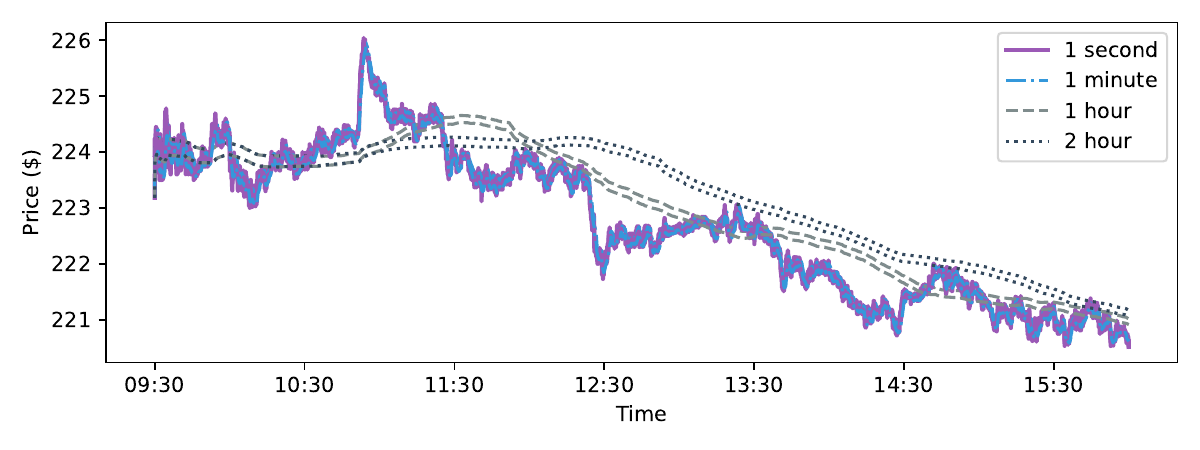} 
    \caption{Rolling averages of AMZN stock's best-bid and best-ask prices across time windows ranging from 1 second to 2 hours. Higher values indicate the best-ask price being above the best-bid price. Notably, sharp changes seen around 11:00 and 12:30 are smoothed in the 1- and 2-hour windows, where shorter-term fluctuations become less prominent.}
    \label{fig:time_resolution_diff}
\end{figure}

\begin{figure} 
    \centering
    \includegraphics[width=\textwidth]{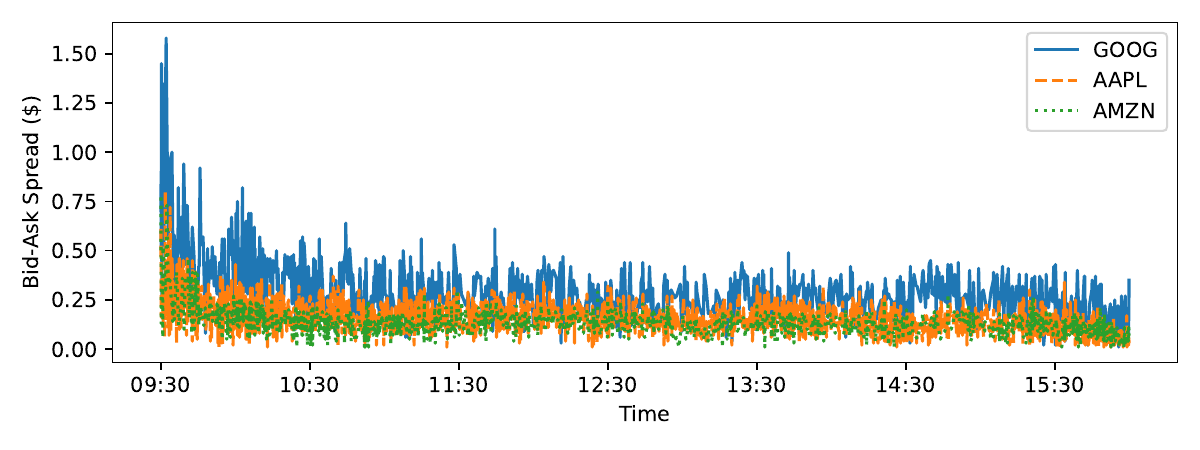} 
    \caption{Bid-Ask Spread (BAS) for a selected trading day, calculated for three sample stocks: GOOG, AAPL, and AMZN. }
    \label{fig:three_bas}
\end{figure}

\subsubsection{High-frequency LOB}
In high-frequency trading and market microstructure, traditional financial assumptions, such as stationarity and independent increments, are often considered inadequate by the real-time, high-frequency nature of the LOB. Unlike in conventional financial modeling, where these assumptions may hold over longer time horizons, the continuous flow of orders, frequent price changes, and rapid reactions by algorithmic trading result in time series data that are inherently noisy, irregular, and non-stationary. This volatility presents unique challenges in modeling and forecasting, as the traditional tools often used in lower-frequency financial data analysis do not sufficiently capture the nuanced behaviors inherent in high-frequency environments.

Figure~\ref{fig:time_resolution_diff} visually depicts these dynamics by showing rolling averages of AMZN stock’s best-bid and best-ask prices over various time intervals—from seconds to hours—within a single trading day. While shorter time windows capture sharper price fluctuations and greater volatility, reflecting the underlying high-frequency dynamics, longer windows smooth out these short-term variations. This visual distinction underscores how time resolution significantly impacts our understanding of data. The choice of time scale in analyzing LOB data is crucial, as high-frequency behaviors may be masked in more aggregated views, potentially overlooking the critical, rapid adjustments inherent to LOB activity.

Further complicating the analysis of LOB data is the often-used assumption of a constant bid-ask spread. This simplification is commonly applied to streamline modeling efforts; however, Figure~\ref{fig:three_bas} demonstrates that the bid-ask spread is far from constant in high-frequency trading. At smaller time intervals, spreads exhibit continuous and significant fluctuations, with notable changes occurring around critical times, such as market open and close. These time-dependent variations in both spread mean and volatility highlight the dynamic complexity of the LOB that a constant spread assumption cannot fully capture. While constant spread models offer mathematical tractability, they fail to represent the richness of real-time market dynamics, especially when considering the effects of microstructure in high-frequency settings \citep{abergel2016limit}.

\subsubsection{Spatiotemporal structure of LOB}

\begin{figure} 
    \centering
    \includegraphics[width=\textwidth]{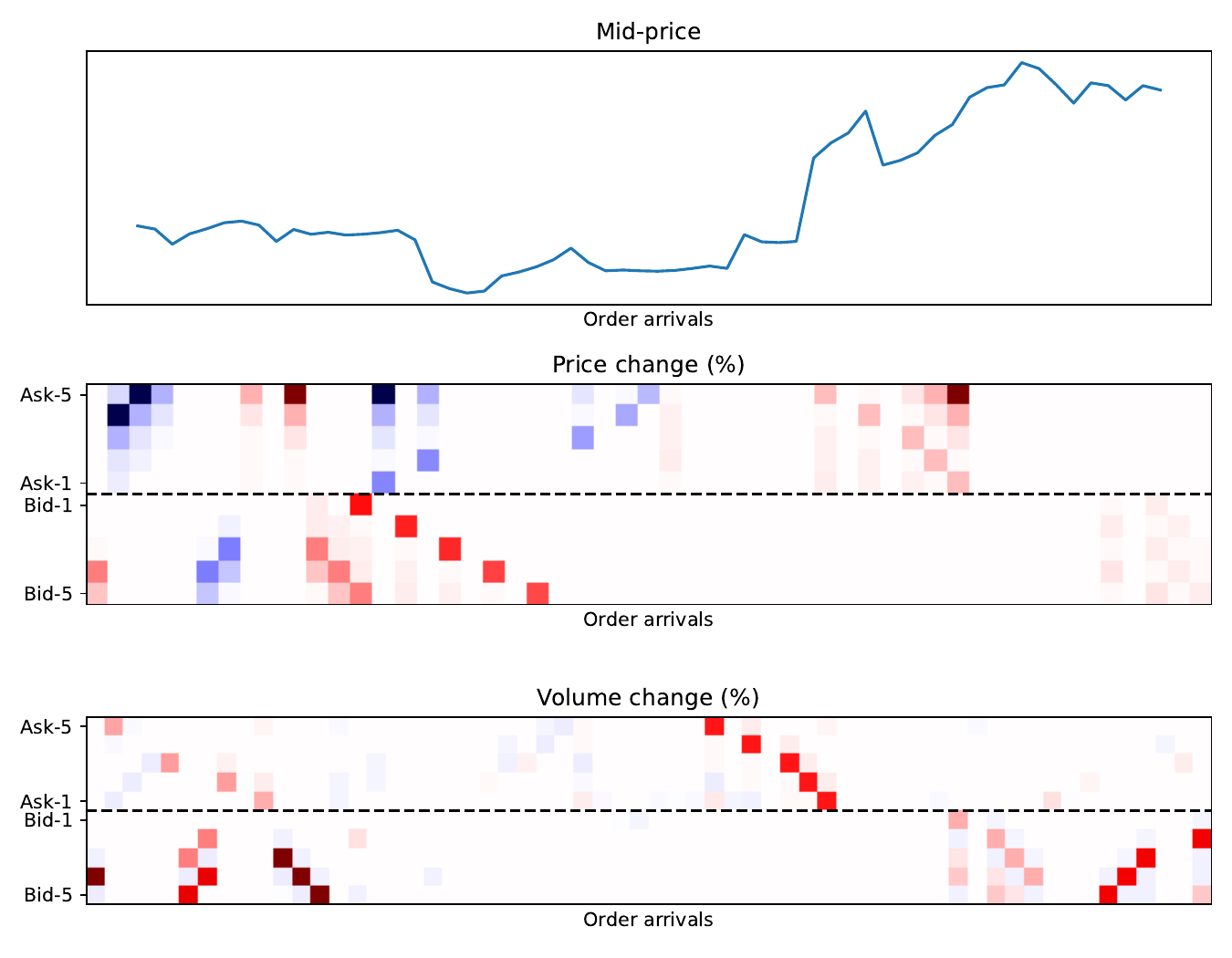} 
    \caption{Mid-price (top), level-5 bid and ask price changes (middle), and level-5 bid and ask volume changes (bottom) aligned along the same order arrivals axis. In the middle and bottom heatmaps, darker shades indicate higher absolute percentage changes, with red representing upward movements and blue representing downward movements; White areas show no change. The mid-price is shown on a raw price scale, while percentage change values are scaled for better comparability across different scales. These plots illustrate data from the first 60 order arrivals of AAPL stock at market open.}
    \label{fig:sparse_changes}
\end{figure}

The high-frequency nature of LOB reveals complex dynamics that influence the order book movements, with multiple components contributing to these fluctuations. In a multi-level LOB, each price level and corresponding volume plays an essential role as new orders are placed, canceled, or executed, reshaping the order book's structure and potentially impacting market behavior across different levels. These rapid and simultaneous changes reveal how the continuous interaction of market participants dynamically impacts both prices and volumes.

Figure~\ref{fig:sparse_changes} underscores this spatiotemporal relationship, presenting how mid-price movements (top panel) are shaped by changes in bid and ask prices and volumes of multiple levels. Price shifts in the multi-level structure reflect the cumulative effect of variations in both price and volume, which tend to be interconnected. In the middle and bottom heatmaps, we observe that large portions of the LOB remain static, showing 0\% change, meaning that not all levels are altered with each order event. This sparse activity, where only specific levels undergo changes, suggests that most of the LOB remains stable between updates, with shifts happening intermittently across some levels. However, when a substantial movement occurs in a certain direction, it often influences neighboring levels, driving future movements in a similar direction. 
Notably, these movements often occur concurrently across various attributes, including price and volume, suggesting a compounded effect across the LOB. This sparse yet correlated activity provides valuable insight into how market depth and order flow dynamically shape the LOB across multiple levels.

These observations emphasize the need for a modeling approach that can accommodate the high-frequency, time-varying, and spatiotemporal properties of LOB data. In this light, a data-driven method, rather than relying on traditional theoretical assumptions, is better suited to capture dependencies that evolve over short intervals. Adopting such an approach allows for a more accurate representation of the dynamics in the LOB, providing insights into market behavior that can inform effective decision-making in high-frequency trading contexts.

\subsection{Data structure and preparation}

We utilize the LOBSTER dataset \citep{huang2011lobster}, which offers real-world LOB data, including both message and order book information for selected stocks. We specifically focus on five major technology companies: AAPL, GOOG, INTC, MSFT, and AMZN. The sample covers the trading day of June 21, 2012, from the market opening at 9:30 am to the closing at 4:00 pm, capturing between 300,000 to 600,000 snapshots per stock, all recorded with millisecond time resolution at Level-5 of the order book. The order book file tracks the evolution of the LOB, while the message file logs the events that trigger updates, with timestamps ranging from milliseconds to nanoseconds.

The order book consists of two types of orders: bids (to buy) and asks (to sell). An LOB is continuously updated with every change in these orders, including new submissions, cancellations, deletions, or executions.
Each snapshot of the LOB data is represented as $\Vec{X_i}$ for $i \in \{1, \dots, I\}$. Both bid and ask orders are captured as vectors containing the timestamp and price-volume pairs for each level $k \in {1,\dots,K}$. Specifically, for the $i^\text{th}$ snapshot, $\Vec{X_i} = \left[ t_i, (p_{1i}^b, v_{1i}^b), \dots, (p_{Ki}^b, v_{Ki}^b), (p_{1i}^a, v_{1i}^a), \dots, (p_{Ki}^a, v_{Ki}^a) \right]$, where $t_i$ is the timestamp, $p_{ki}^{b/a}$ represents the level-$k$ bid/ask price, and $v_{ki}^{b/a}$ represents the level-$k$ bid/ask volume. To maintain consistency, we fix the number of levels, $K$, at 5 for both bid and ask orders in our experiments.

To ensure consistency and facilitate analysis, we standardized the timestamps to equal intervals of 5 seconds during data cleansing. For example, this results in approximately 4,680 snapshots per trading day. We also concatenated the data from all five stocks, creating a dataset with 100 dimensions, including five levels of bid and ask prices and volumes for each stock.
The dataset was divided into training, validation, and testing sequences in a 6:2:2 ratio. For prediction, a context sequence of 120 time steps (equivalent to 10 minutes) was used to output a target sequence of 24 time steps (equivalent to 2 minutes).

\subsection{Input transformation and scaling}

% \paragraph{Stationary scaling transformation}

\begin{figure} 
    \centering
    \includegraphics[width=.9\textwidth]{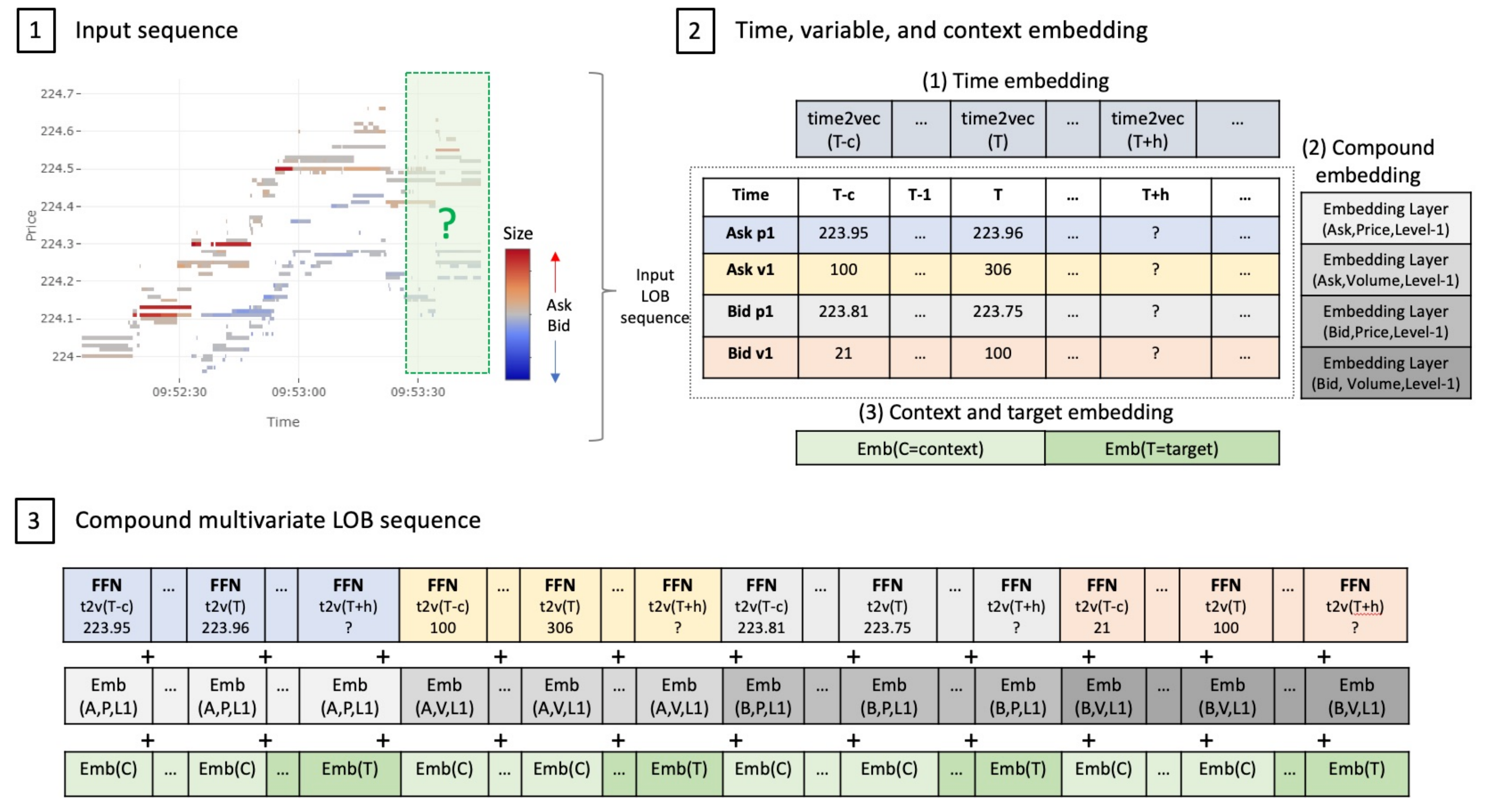} 
    \caption{Input encoding pipeline for limit order book forecasting. (1) Level-5 LOB sequences are split into 10-minute context points and 2-minute target points. (2) Time, variable, and context embeddings are created to structure the input sequence. (3) The final LOB input sequence is generated by integrating these embedding layers.}
    \label{fig:lob_reader}
\end{figure}

The input encoding pipeline for LOB prediction involves pre-processing the LOB data to be fed into a forecasting model (Figure~\ref{fig:lob_reader}). This step involves extracting relevant features from the LOB data, such as the price, volume, and time of each order. However, non-stationary dimensions in the data can hinder learning by causing distributional shifts between the training and testing periods, a common challenge when dealing with long sequences of time series data. For example, multi-level prices often exhibit non-stationarity even within a single trading day.

\begin{figure} 
    \centering
    \includegraphics[width=.49\linewidth]{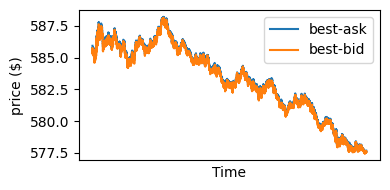} 
    \includegraphics[width=.49\linewidth]{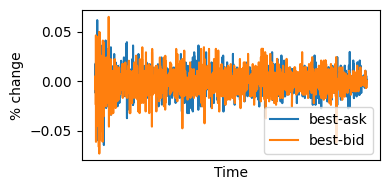}  
    \caption{Price movement of a limit order book: raw time series (left) and percent-change transformed time series (right).}
    \label{fig:series_transform}
\end{figure}

In our experiment, we employ percent-change transformation of prices to facilitate training. The percent change transformation is applied to multivariate price sequences at all levels using the following equation:
\begin{equation}
     {p}_{k,i}^{\text{perc}} = \frac{ {p}_{k,i} -  {p}_{k,i-1}}{ {p}_{k,i-1}} 
    \label{eq:percent_transform}
\end{equation}
Figure~\ref{fig:series_transform} illustrates that the transformed series exhibits greater stationarity in price movements compared to the raw series.
During training, we also apply min-max scaling to price and volume on a variable-wise basis to ensure that variables with large fluctuations do not disproportionately influence the model. The min-max scaling transformation for price and volume is defined as: 
\begin{equation}
 {p}_{k,i}^{\text{scaled}} = \frac{ {p}_{k,i}^{\text{perc}} - \min_i(  {p}_{k,i}^{\text{perc}})}{\max_i( {p}_{k,i}^{\text{perc}}) - \min_i(  {p}_{k,i}^{\text{perc}})}, \quad 
 {v}_{k,i}^{\text{scaled}} = \frac{ {v}_{k,i} - \min_i(  {v}_{k,i} )}{\max_i( {v}_{k,i}) - \min_i(  {v}_{k,i} )}
\label{eq:minmax}
\end{equation}
The performance of various transformation methods is compared in a later section, specifically in Table~\ref{tab:performance_input} within Section~\ref{preliminary}.

\section{Method}

\subsection{Compound multivariate embedding}

The embedding process encodes time, values, variables, and context-target information. We employ the Time2Vec layer \citep{kazemiTime2VecLearningVector2019} to transform timestamps into frequency embeddings, which are then passed through a feed-forward layer to capture temporal patterns and data sequence. Additionally, a binary context-target embedding identifies whether a value is from the context or is to be forecasted, enabling the model to distinguish between observed and future data points.

\begin{figure} 
    \centering
    \includegraphics[width=\textwidth]{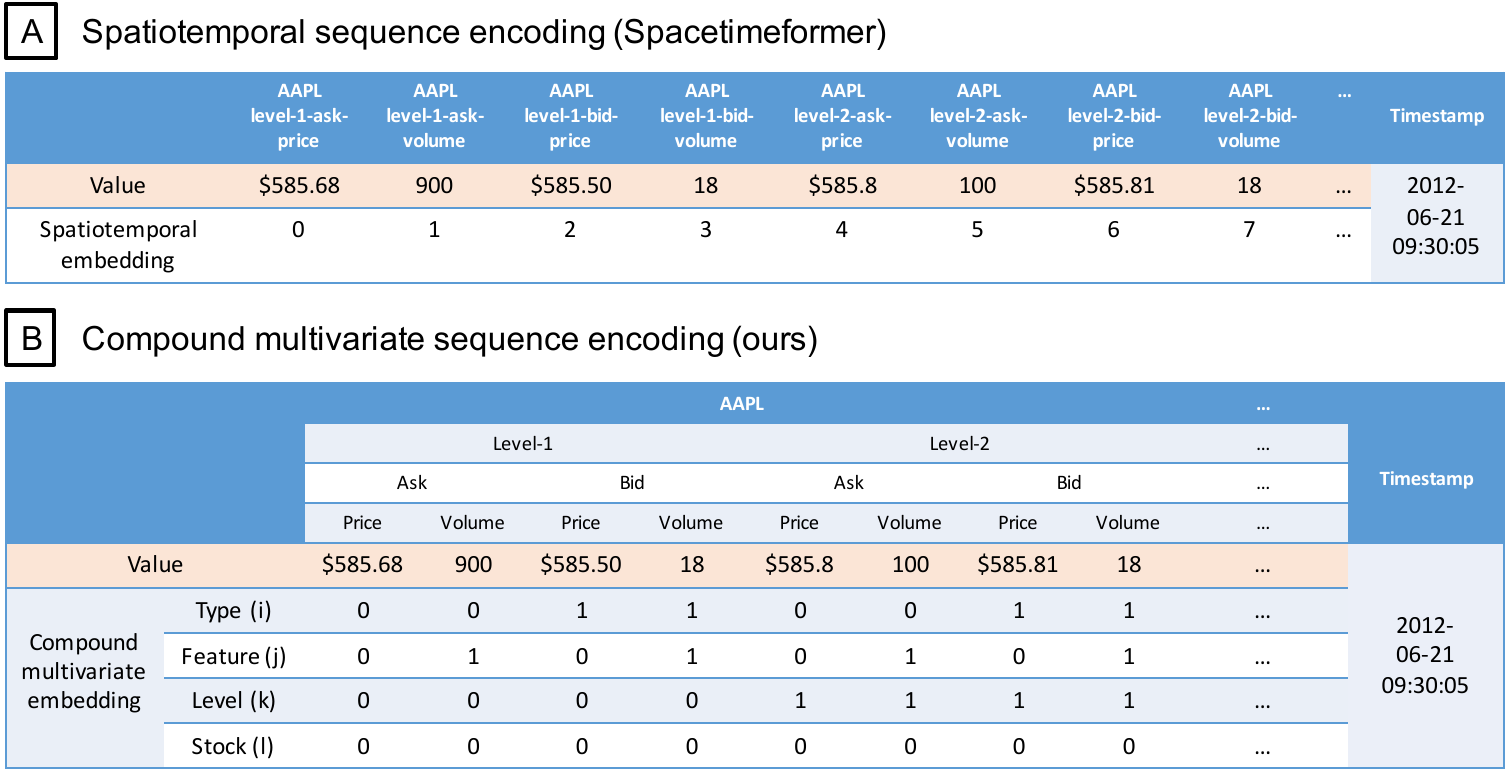} 
    \caption{Input encoding pipeline for limit order books: spatiotemporal (A) and compound multivariate (B) embedding.}
    \label{fig:embedding_method}
\end{figure}

Our data structure incorporates compound attributes such as multiple levels (level-$k$), types (bid or ask), features (price or volume), and stocks (tickers). To effectively capture dependencies among these attributes, we developed a compound multivariate embedding method inspired by the spacetimeformer approach \citep{grigsbyLongRangeTransformersDynamic2022}. Unlike the original spacetimeformer, which assigns embedding integers to each variable separately, our method assigns embeddings to each attribute. Then, we combine and scale the multi-level embeddings. As illustrated in Figure~\ref{fig:embedding_method}, each attribute is represented by its own embedding layer. For instance, one layer might encode whether the data is a bid or ask, another might encode the order level (e.g., level-1, level-2), and another might encode the feature type (price or volume).

\begin{figure} 
    \centering
    \includegraphics[width=\textwidth]{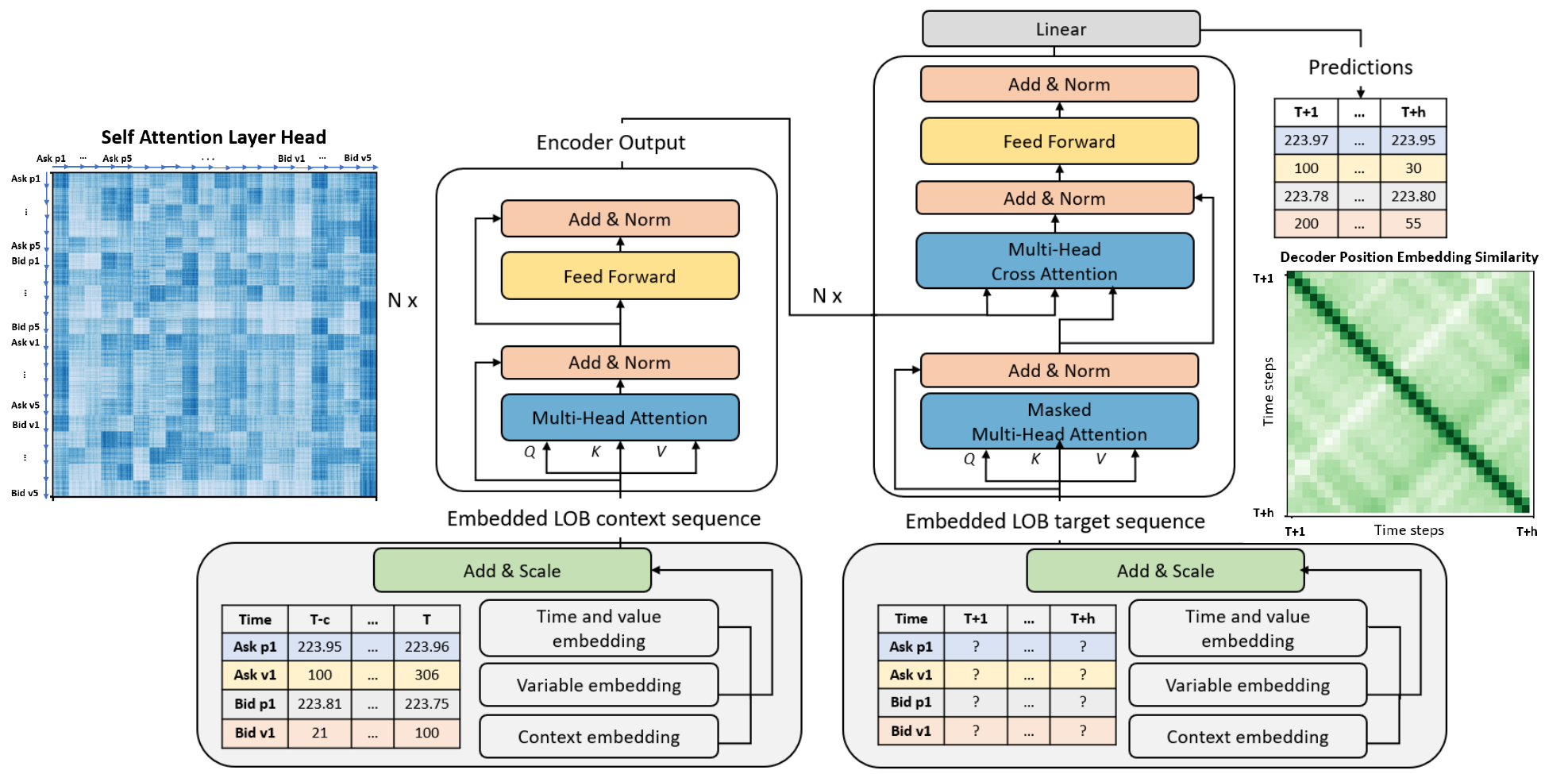} 
    \caption{The Spatiotemporal attention architecture applied to the sequence of limit order books. The darker shades indicate higher attention scores. This approach is based on the work of \citet{grigsbyLongRangeTransformersDynamic2022}.}
    \label{fig:lob_architecture}
\end{figure}

By integrating and scaling these multiple embedding layers, our approach captures not only temporal and spatial patterns but also the complex relationships between different attributes within the LOB data. This method also reduces the number of parameters in the embedding layer while effectively incorporating interdependencies across attributes. The embedded sequence is processed using an attention-based encoder-decoder architecture (Figure~\ref{fig:lob_architecture}). The encoder employs self-attention modules to capture both local and global contexts, while the decoder uses masked attention to prevent access to future information during prediction. The final feed-forward layer transforms the output back into the original LOB sequence format.

The attention mechanism \citep{vaswaniAttentionAllYoua} updates token representations and re-normalizes the resulting matrix: 
\begin{equation} \text{Attention}(Q,K,V) \in \mathbb{R}^{L_x \times d} = \text{softmax} \left( \frac{Q K^T}{\sqrt{d}}\right) V, \end{equation} 
where query vectors $Q=W^Q X$, key vectors $K=W^K Z$, and value vectors $V=W^V Z$ are derived from learned parameters. For a sequence of length $L$ and dimension $d$, the self-attention matrix outputs an $(L \times d)^2$ matrix, enabling assessment of both temporal and inter-variable dependencies.
In our experiment, we use the Performer method \citep{choromanski2020rethinking} to enhance computational efficiency, which approximates attention with linear space and time complexity.

\subsection{Training}

\paragraph{Forecasting loss}
\label{sec:evaluation}

For our evaluation, we measure the difference between predicted and actual limit orders using mean squared error (MSE) and mean absolute error (MAE) as our primary metrics. The forecasting loss specifically focuses on minimizing the MSE of level-5 price and volume predictions within our sample dataset:
\begin{align}
    \text{Forecasting Loss} = \sum_{i,j,k} ({p}_{k,i}^j- \hat{p}_{k,i}^j)^2 + \sum_{i,j,k} ({v}_{k,i}^j- \hat{v}_{k,i}^j)^2.
\end{align}
MSE is chosen because of its effectiveness in addressing multivariate forecasting tasks, particularly in transformer-based models, and its ability to reduce significant discrepancies in predictions. Although metrics like MAE are recorded for reference, they are not used as loss functions.

\paragraph{Structural regularizer}

To ensure that our model incorporates the multi-level limit order book (LOB) structures during the training process, we introduce a structural regularizer. For instance, the level-1 ask (bid) price is defined as the lowest (highest) offer price available from sellers (buyers) quoting a security, while level-2 prices represent the second-best offers, and so forth. Let $p_{ki}^a$ ($p_{ki}^b$) represent the level-$k$ ask (bid) price for the $i$-th snapshot, where $k$ denotes the order level. For any fixed $i$-th LOB snapshot, the multi-level price structure is given by:
\begin{align}
p_{k_1 i}^a < p_{k_2 i}^a &\quad \text{for all } k_1 < k_2, \\
p_{k_1 i}^b < p_{k_2 i}^a &\quad \text{for } k_1 = k_2 = 1, \\
p_{k_1 i}^b > p_{k_2 i}^b &\quad \text{for all } k_1 < k_2
\end{align}
To nudge our network outputs adhere to these ordinal price structures, we add a regularizing term to capture this restriction at each snapshot-$i$:
\begin{align}
\text{Structure loss}_i =  \sum_{k=1}^{K-1} \left( \text{ReLU}(\hat{p}_{k,i}^a - \hat{p}_{k+1,i}^a) + \text{ReLU}(\hat{p}_{k+1,i}^b - \hat{p}_{k,i}^b) \right) + \text{ReLU}(\hat{p}_{1,i}^b - \hat{p}_{1,i}^a), 
\end{align} 
where $p_{k,i}^j$ refers to the actual price, $\hat{p}_{k,i}^j$ refers to the predicted price, and the ReLU (Rectified Linear Unit) function is defined as $\text{ReLU}(x) = \max(0, x)$. This function ensures that only positive differences contribute to the penalty, thus enforcing the ordinal structure by penalizing any violation where the predicted lower-level ask price is not less than the higher-level ask price, or where the predicted lower-level bid price is not greater than the higher-level bid price. 

The total loss function is defined with a regularization weight $w_o \geq 0$, which determines the strength of the regularization:
\begin{align}
    \text{Loss} =  \text{Forecasting loss}+ w_o \cdot \sum_{i=1}^I \text{Structure loss}_i,
\end{align}
This function penalizes predictions that break the ordinal structure of price levels while also minimizing the forecasting error, thereby ensuring that the correct multi-level sequences are maintained during training. We fixed $w_o$ at 0.01 in our experiment for balancing the scale of forecasting and structure losses.

For training, the learning rate decay is managed with a decay factor of 0.8 and 1000 warmup steps. Additionally, the model employed seasonal decomposition and reversible normalization \citep{kim2021reversible} techniques to enhance performance. We use three-head attention modules for training the attention-based models. We monitored the validation loss across epochs and halted the process if no improvement was observed after 10 consecutive epochs. For testing, we selected the trained model weights associated with the lowest validation total loss. The training was conducted on Google Colab using a single A100 GPU.

% Figure~\ref{fig:AAPL_forecast} illustrates the effect of the ordinal structure regularizer on the predicted sequences. The left panel, trained without the ordinal loss (equivalent to $w_o=0$), exhibits some violations of the level structure. In contrast, the right panel, trained with the ordinal loss ($w_o>0$), preserves the ordinal structure consistently from level-1 onward.

% \begin{figure} 
%     \centering
%     \includegraphics[width=.49\textwidth]{lob/AAPL_forecast_none.png} 
%     \includegraphics[width=.49\textwidth]{lob/AAPL_forecast_ordinal.png}  
%     \caption{Price forecasting of a limit order book snapshot: model trained without ordinal loss (left) and model trained with ordinal loss (right)}
%     \label{fig:AAPL_forecast}
% \end{figure}

\section{Results}
\label{preliminary}

% \usepackage{booktabs}
% \usepackage{multirow}
\begin{table}[]
\centering
\begin{threeparttable}
\caption{Performance metrics calculated for each input transformation calculated for each input transformation across the top-5 price levels. Mean Squared Error (MSE) and Mean Absolute Error (MAE) are provided for each method. All values have been descaled and are presented in their original units of dollars.}
% \usepackage{booktabs}
% \usepackage{multirow}
% \usepackage[normalem]{ulem}
\begin{tabular}{@{}cccc@{}} 
\toprule
Input transformation & Percent-change & Min-max & Percent-change + Minmax \\ \midrule
MSE                  & 4.5447         & 0.0660  &  \textbf{0.0464}             \\
MAE                  & 1.3225         & 0.1469  &   \textbf{0.1260}          \\ \bottomrule
\end{tabular} 
% \begin{tablenotes} 
% \end{tablenotes}
\label{tab:performance_input}
\end{threeparttable}
\end{table}

\begin{table}[]
\centering
\begin{threeparttable}
\caption{Prediction results for sample test data: Mean Squared Error (MSE) and Mean Absolute Error (MAE) are computed for each method across selected dimensions. The mid-price is determined by averaging the best bid and ask prices. Loss metrics are provided for mid-price, price, and volume on level-5 limit order book (LOB) data.  Forecasting loss refers to MSE losses from all five levels of prices and volumes, while structure loss quantifies the violations of the ordinal structure. Total loss represents the weighted sum of the forecasting loss and structure loss. All values are calculated after scaling.} 
\begin{tabular}{@{}cccc|ccc@{}}
\toprule
\multicolumn{1}{l}{}       & \multicolumn{1}{l}{}                                         & \multicolumn{2}{l}{Time series models}                   & \multicolumn{3}{c}{Attention-based models}                                                                                                                      \\ 
% \midrule
\cmidrule(r){3-4}\cmidrule(l){5-7} 
                           & \multicolumn{1}{c}{ }  & \multicolumn{1}{c}{Linear} & \multicolumn{1}{c}{LSTM} & \multicolumn{1}{c}{Temporal} & \multicolumn{1}{c}{Spacetime} & 
                           \multicolumn{1}{c}{Compound (ours)} \\ \midrule
\multirow{2}{*}{mid-price} & MSE                                                          & 0.0125                        &  0.0124             & 0.0124                       &  0.0122                              &   \textbf{0.0120}                                                                                    \\
                           & MAE                                                          & 0.1501                        & 0.1466                 & 0.1520                       & 0.1425                              &  \textbf{0.1388}                                                                              \\
\multirow{2}{*}{price\tnote{*}}     & MSE                                                          & 0.0026                        & 0.0026                  & 0.0027                       &  \textbf{ 0.0025}                              &   \textbf{0.0025}                                                                               \\
                           & MAE                                                          & 0.0310                        & 0.0303                & 0.0332                       & 0.0293                              &   \textbf{0.0288}                                                                             \\
\multirow{2}{*}{volume\tnote{**}}    & MSE                                                          &  \textbf{0.0105}                        &  0.0118              & 0.0109                       &  \textbf{0.0105}                             & 0.0106                                                                                     \\
                           & MAE                                                          & 0.0510                        &  0.0541             & 0.0533                       & 0.0515                              &  \textbf{0.0504}                                                                                      \\
\multicolumn{2}{c}{Forecasting loss}                                                          &  \textbf{0.0065}                    &  0.0072                    & 0.0068                       &  \textbf{0.0065}                          &   0.0066                                                                             \\
                         %  & MAE                                                          & 0.0410                        & {\ul 0.0390}             & 0.0433                       & 0.0405                              & 0.0396                                                                                     \\
\multicolumn{2}{c}{Structure loss}                                                       & 0.2430                        & 0.2624                   & 0.8836                       & 0.5774                              &   \textbf{0.1480}                                                                              \\ \hline
\multicolumn{2}{c}{Total loss }                                   & 0.0090                        & 0.0098                   & 0.0157                       & 0.0123                              & \textbf{0.0080}                                                                               \\ \bottomrule
\end{tabular}
\begin{tablenotes}
\scriptsize
\item[*] Price metrics refer to the aggregated loss for bid and ask prices across all five levels.
\item[**] Volume metrics refer to the aggregated loss for bid and ask volumes across all five levels.
% \item[***] Entire refers to the sum of losses from all five levels of prices and volumes.
\end{tablenotes}
\label{tab:performance}
\end{threeparttable}
\end{table}

To assess the impact of input transformations on prediction accuracy, we first examine the results in Table~\ref{tab:performance_input}. Mean Squared Error (MSE) and Mean Absolute Error (MAE) metrics are reported for different transformations applied to multi-level price data. The combined approach, using both percent-change and min-max scaling, achieves the lowest error rates with an MSE of 0.0464 and an MAE of 0.1260, outperforming each transformation used individually. These values are presented in the raw scale to allow a consistent comparison of transformation effects.

Percent-change normalization, a standard transformation in financial time series, enforces stationarity in input sequences—an essential condition in time series forecasting. In machine learning, non-stationarity of time series can lead to unreliable predictions, as trends or shifts may skew the model's focus away from the underlying fluctuations. By transforming prices into percent changes, the model can emphasize relative movements around a baseline, capturing meaningful patterns without the distraction of directional trends. 

Min-max scaling, on the other hand, is well-suited for multivariate data where different scales are present, as in order volume and price levels. Volume data often has higher variance than price data, and min-max scaling helps optimize learning across price levels by constraining these variations. This approach prevents extreme volume values from overwhelming the model’s attention, enabling it to respond effectively to both price and volume information.

Our experimental results demonstrate that combining percent-change normalization and min-max scaling maximizes the benefits of each transformation. By stabilizing the price series for stationarity and managing volume variations within a fixed range, this combined approach achieves more accurate and consistent predictions across multi-level price data, resulting in enhanced model performance and reliability.

We now present the prediction results for test data, with a focus on various performance metrics discussed in Section~\ref{sec:evaluation}. Our experiments compare various sequence-to-sequence prediction models of prior work, including a linear autoregressive model (\textit{Linear}), a standard LSTM network \citep{schmidhuber1997long} (\textit{LSTM}), a transformer with temporal attention (\textit{Temporal}) as introduced in Informer \citep{zhou2021informer}, and a spacetimeformer model (\textit{Spacetime}) that incorporates both spatial and temporal attention \citep{grigsbyLongRangeTransformersDynamic2022}. We also evaluate the performance of our attention-based model with compound multivariate embedding (\textit{Compound}) in comparison to the other techniques.

Table~\ref{tab:performance} compares overall prediction results of various models on level-5 LOB data, with the compound attention-based model (ours) showing the best overall performance. This model achieved the lowest error in all price-related metrics, including mid-price and multi-level prices, outperforming both traditional time series models and other attention-based models. Notably, the compound model excelled in maintaining ordinal structure with the lowest structure loss, leading to the lowest total loss, indicating a strong balance of accuracy and structural consistency across mid-price, price, and volume forecasting. This highlights the method's superior ability to capture structural dependencies through the compound multivariate embedding process. Although metrics calculated from mid-price, multi-level prices and volumes are recorded for reference, they are not used as loss functions.

\begin{figure} 
    \centering
    \includegraphics[width=.49\textwidth]{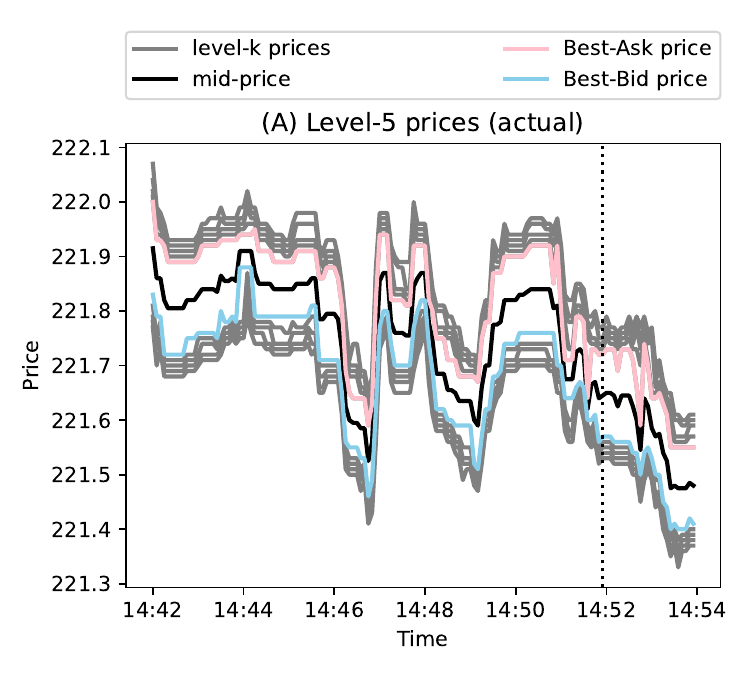} 
    \includegraphics[width=.49\textwidth]{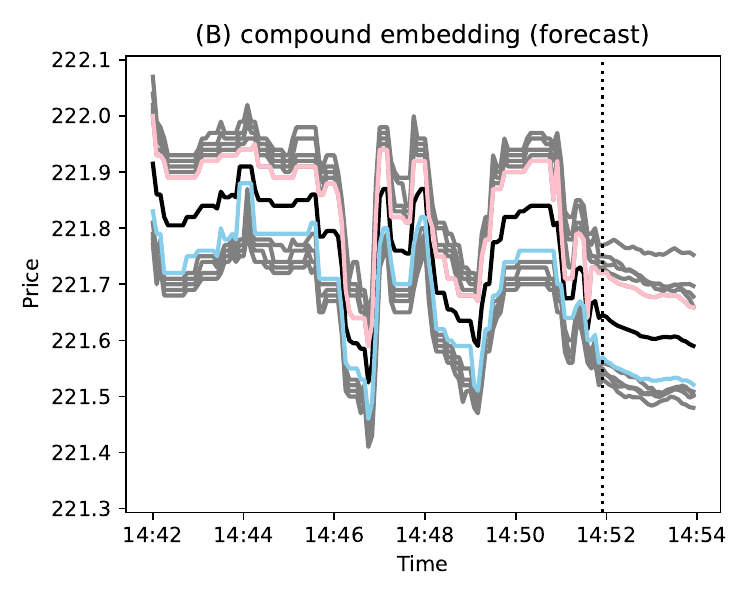} 
    \includegraphics[width=.49\textwidth]{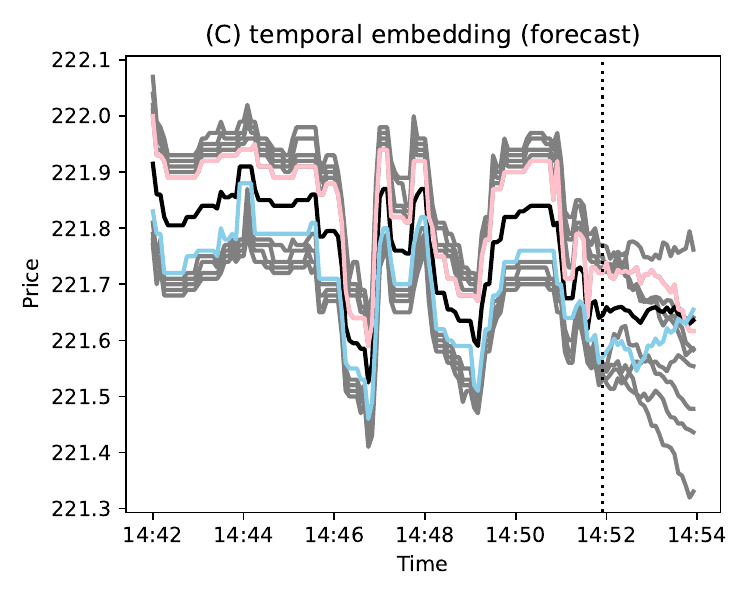} 
    \includegraphics[width=.49\textwidth]{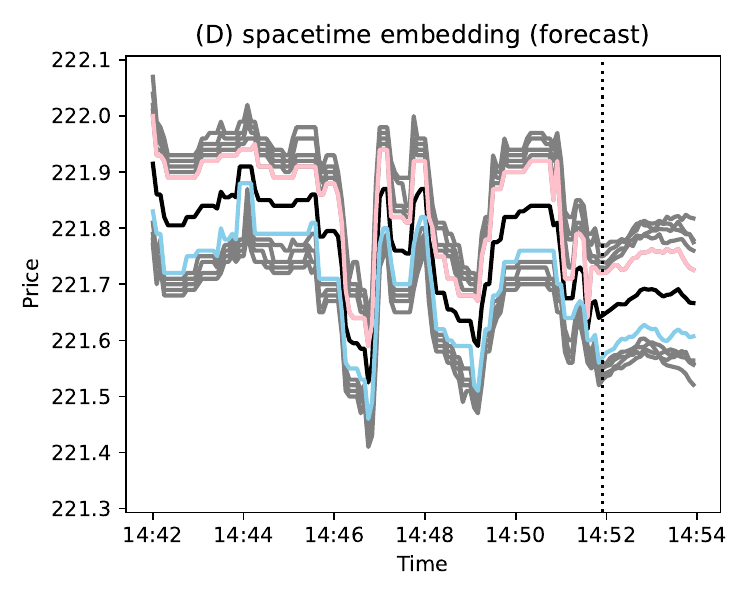} 
    \caption{Price forecasting of a limit order book for AMZN stock: The graphs predict level-5 prices 2 minutes ahead, based on the preceding 10 minutes of context, with the prediction period starting on the right side of the dotted line. Panel A (top left) shows the actual price movements, while panels B, C, and D display the predicted price movements using three attention-based methods: compound multivariate embedding (B), temporal embedding (C), and spatiotemporal embedding (D). Multi-level prices, excluding the best prices, are represented in sequential grey lines.}
    \label{fig:AMZN_example}
\end{figure}

\begin{figure} 
    \centering
     \includegraphics[width=.45\textwidth]{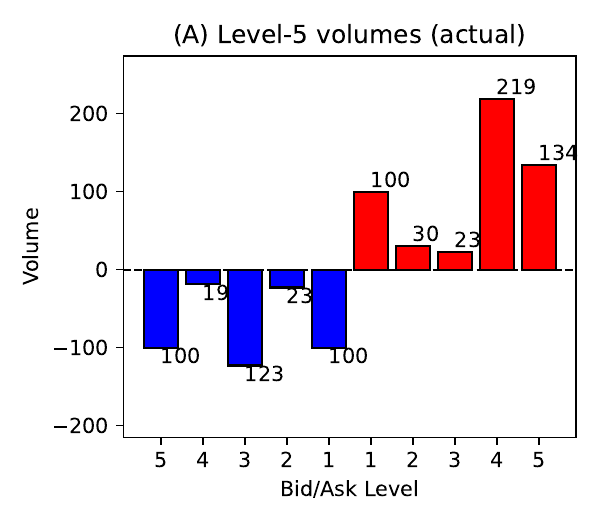} 
    \includegraphics[width=.45\textwidth]{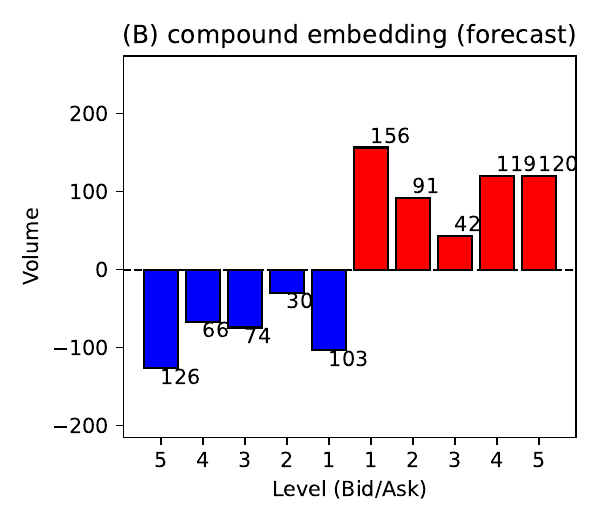} 
    \includegraphics[width=.45\textwidth]{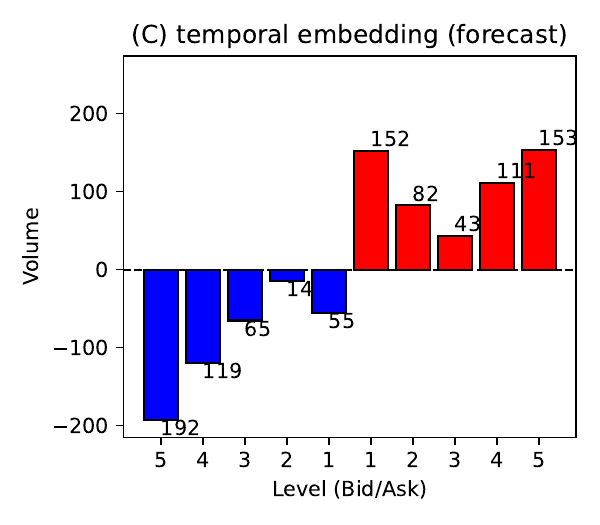} 
    \includegraphics[width=.45\textwidth]{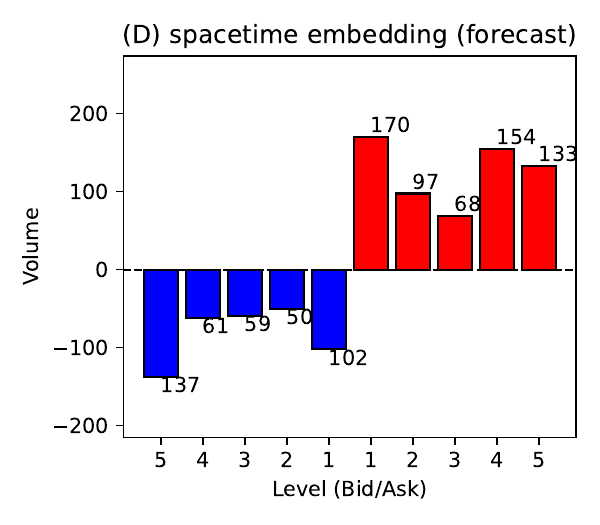} 
    \caption{Volume forecasting of a limit order book snapshot for AMZN stock: The graphs forecast level-5 volumes for the first snapshot of the testing period based on the preceding 10 minutes of context. Panel A shows the actual order sizes, while panels B, C, and D display the predicted volume snapshots using three attention-based methods: compound multivariate embedding (B), temporal embedding (C), and spatiotemporal embedding (D).}
    \label{fig:AMZN_example_vol}
\end{figure}

Figure~\ref{fig:AMZN_example} and Figure~\ref{fig:AMZN_example_vol} provide visual comparisons of the actual versus predicted price and volume movements for the level-5 LOB data of AMZN. We selected the first window of the test period for comparison. In Figure~\ref{fig:AMZN_example}, 
the price forecasting results demonstrate that the compound multivariate embedding method (B) closely tracks the downward trend of actual price movements (A), offering more accurate predictions than the temporal (C) and spatiotemporal (D) embedding methods. While both our approach and the spatiotemporal embedding effectively capture the overall ordinal structure of multi-level prices, the temporal embedding fails to preserve the ordinal relationships between dimensions. This aligns with the concepts of the embedding methods, as the temporal embedding processes multiple time series without incorporating variable embeddings, resulting in structural violations.

\begin{figure} 
    \centering
    \includegraphics[width=.49\textwidth]{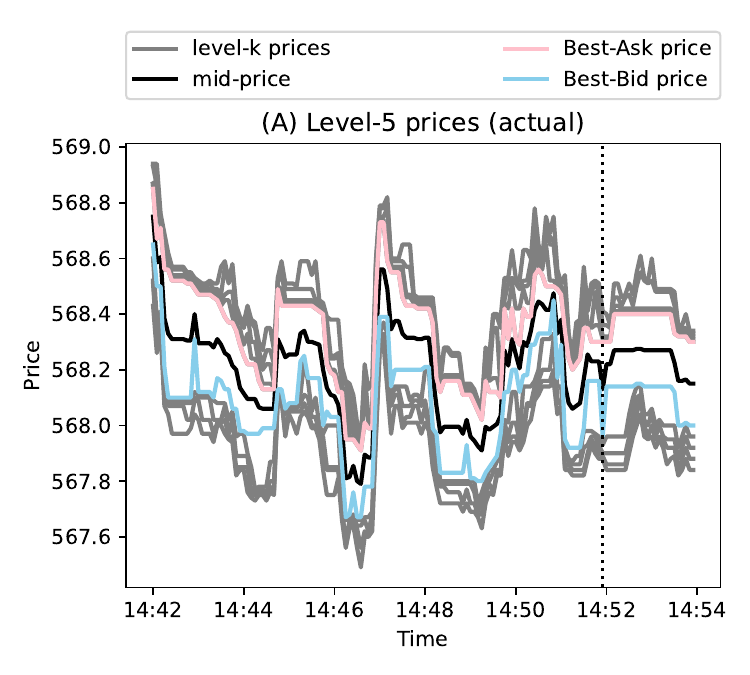} 
    \includegraphics[width=.49\textwidth]{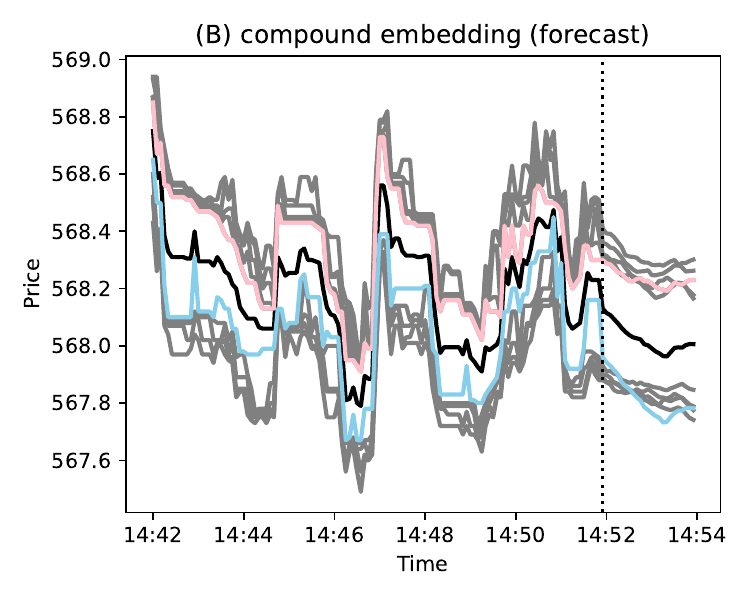} 
    \includegraphics[width=.49\textwidth]{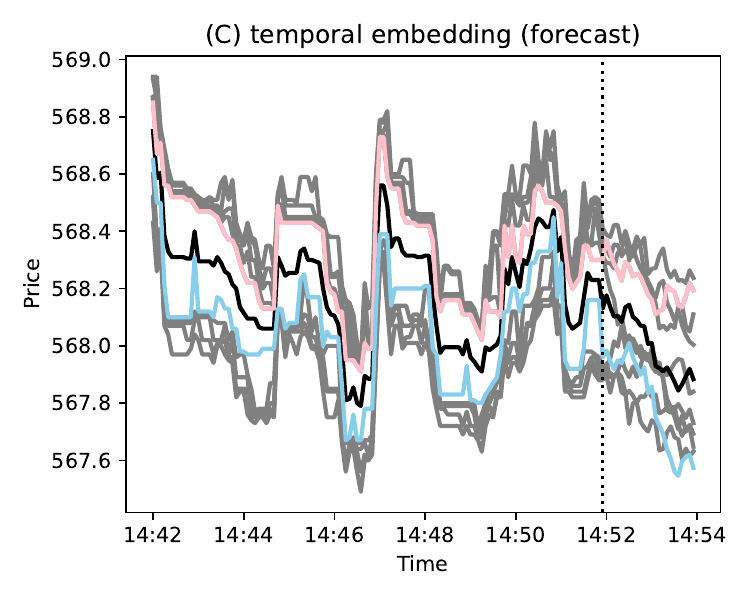} 
    \includegraphics[width=.49\textwidth]{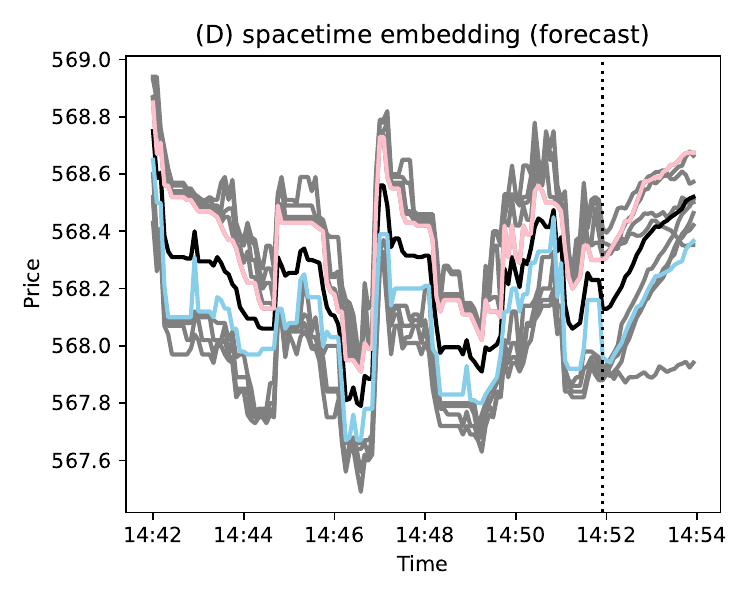} 
     \includegraphics[width=.49\textwidth]{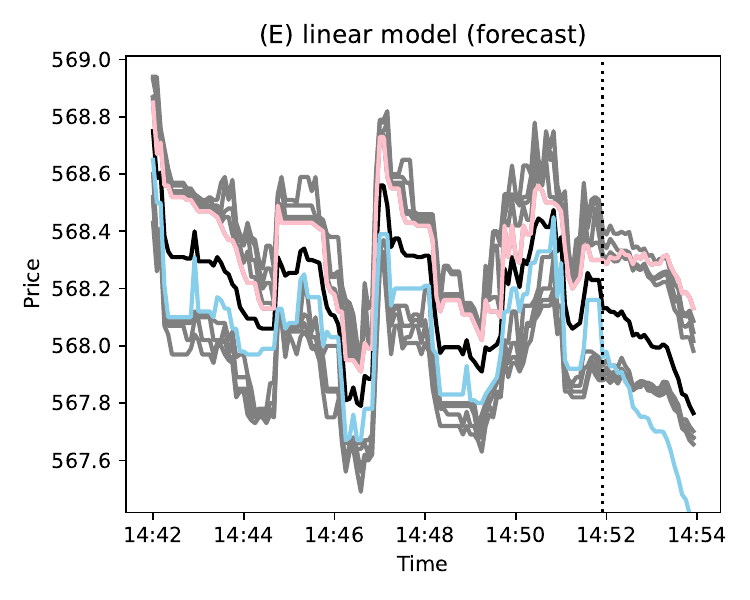} 
    \includegraphics[width=.49\textwidth]{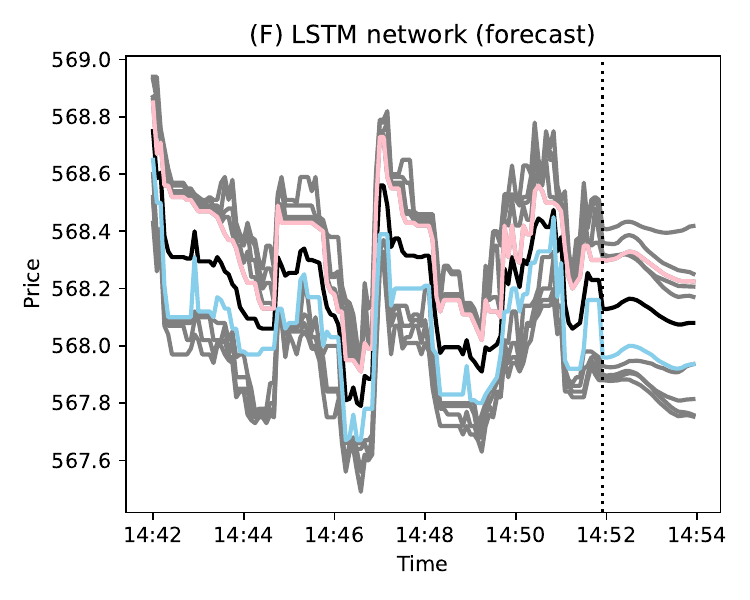} 
    \caption{Price forecasting of a limit order book for GOOG stock: The graphs predict level-5 prices 2 minutes ahead, based on the preceding 10 minutes of context, with the prediction period starting on the right side of the dotted line. Panel A (top left) shows the actual price movements, while panels B, C, and D display the predicted price movements using five forecasting methods: compound multivariate embedding (B), temporal embedding (C), spatiotemporal embedding (D), linear model (E), and LSTM network (F). Multi-level prices, excluding the best prices, are represented in sequential grey lines.}
    \label{fig:GOOG_example}
\end{figure}

\begin{figure} 
    \centering
    \includegraphics[width=.49\textwidth]{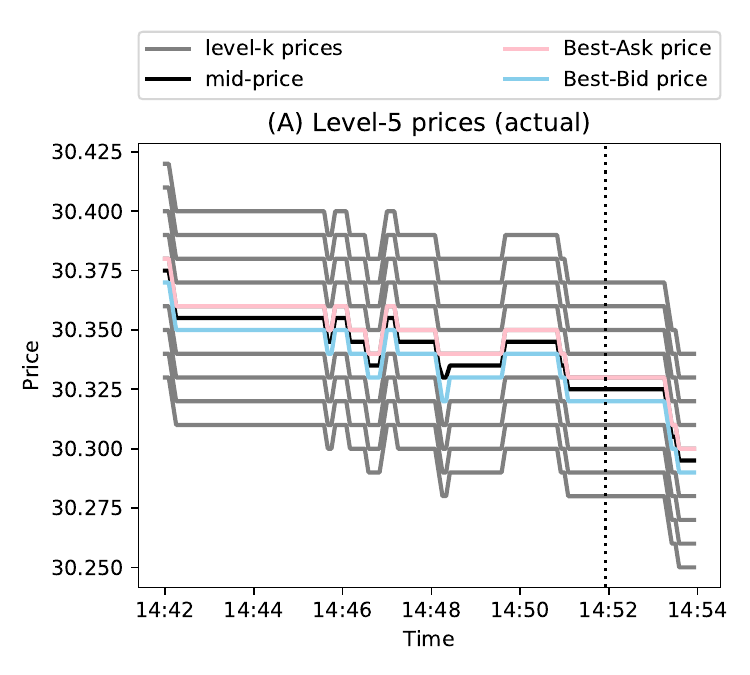} 
    \includegraphics[width=.49\textwidth]{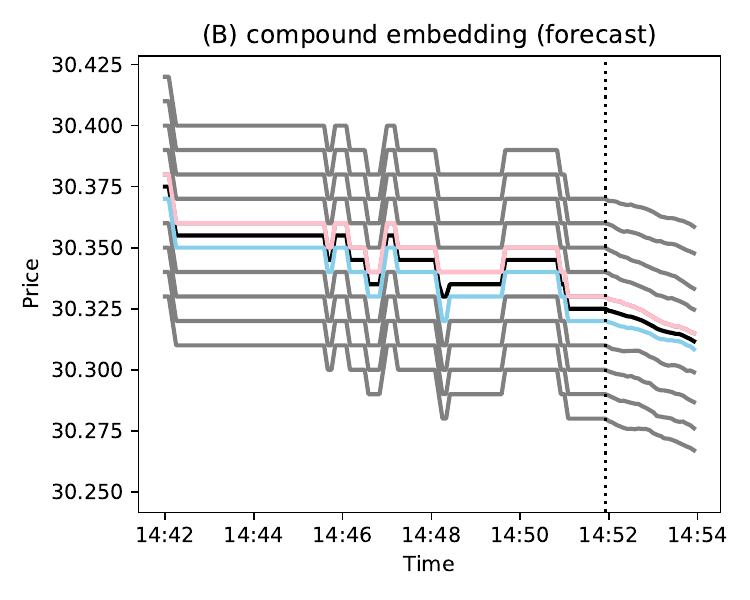} 
    \includegraphics[width=.49\textwidth]{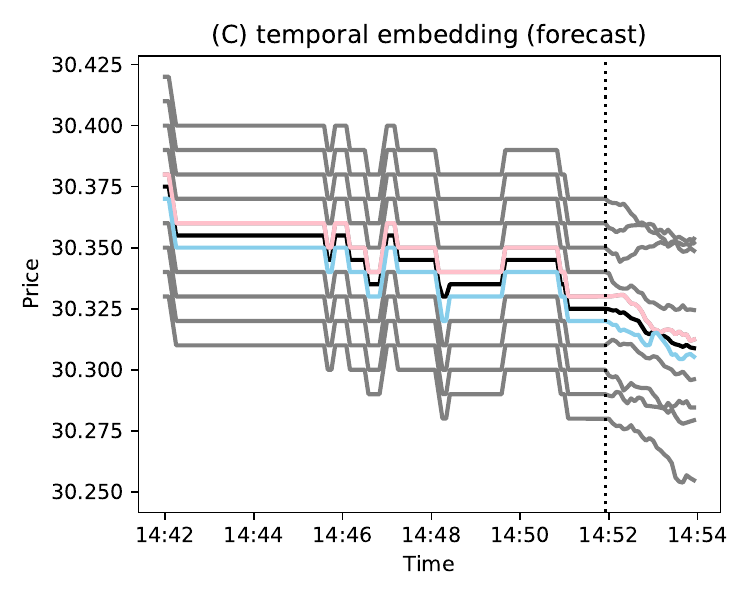} 
    \includegraphics[width=.49\textwidth]{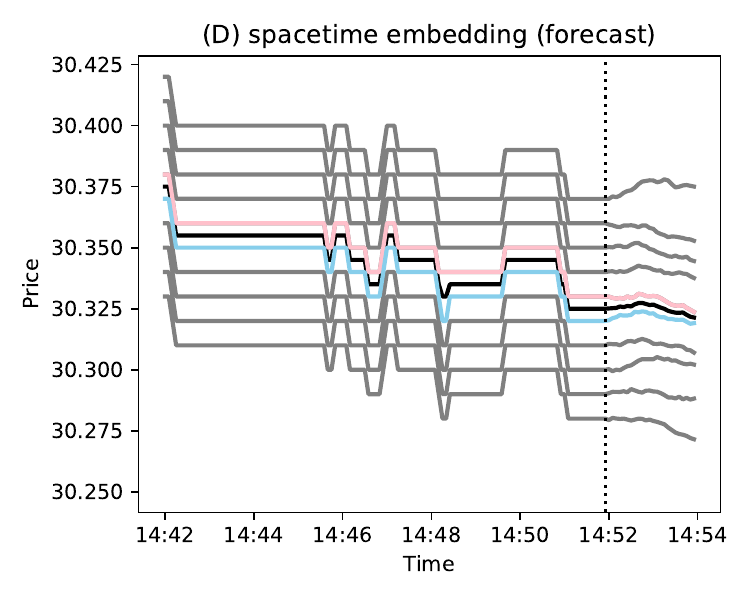} 
    \includegraphics[width=.49\textwidth]{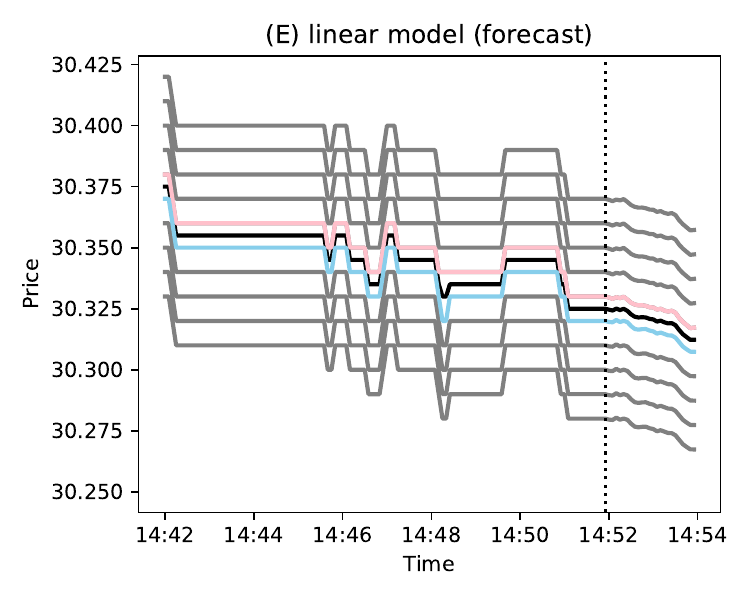} 
    \includegraphics[width=.49\textwidth]{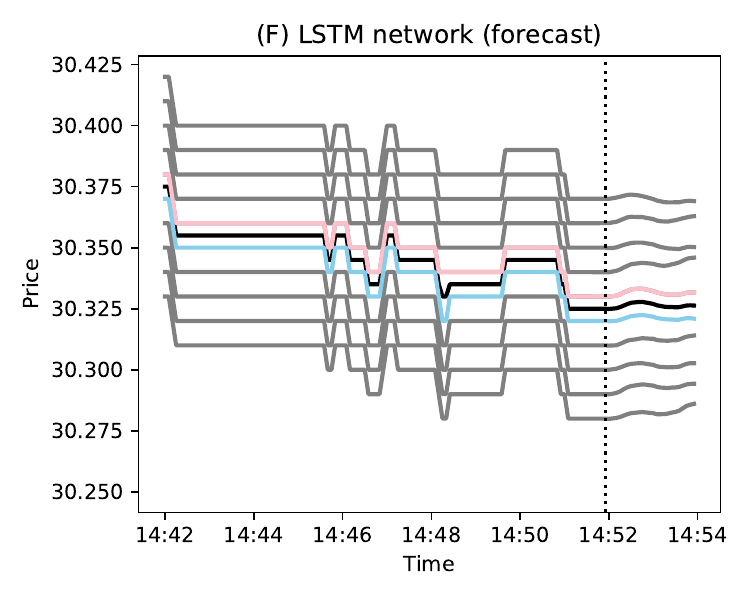} 
    \caption{Price forecasting of a limit order book for MSFT stock: The graphs predict level-5 prices 2 minutes ahead, based on the preceding 10 minutes of context, with the prediction period starting on the right side of the dotted line. Panel A (top left) shows the actual price movements, while panels B, C, and D display the predicted price movements using five forecasting methods: compound multivariate embedding (B), temporal embedding (C), spatiotemporal embedding (D), linear model (E), and LSTM network (F). Multi-level prices, excluding the best prices, are represented in sequential grey lines.}
    \label{fig:MSFT_example}
\end{figure}

\begin{figure} 
    \centering
    \includegraphics[width=.49\textwidth]{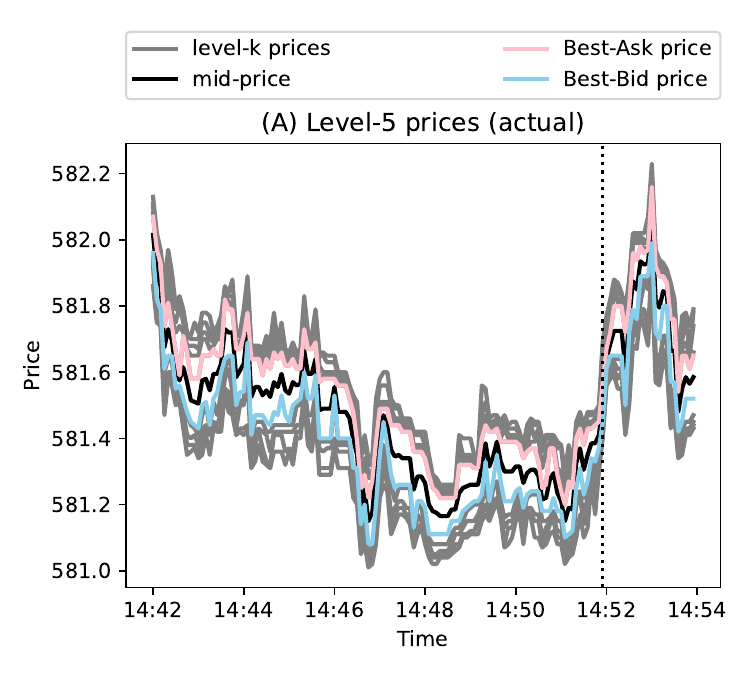} 
    \includegraphics[width=.49\textwidth]{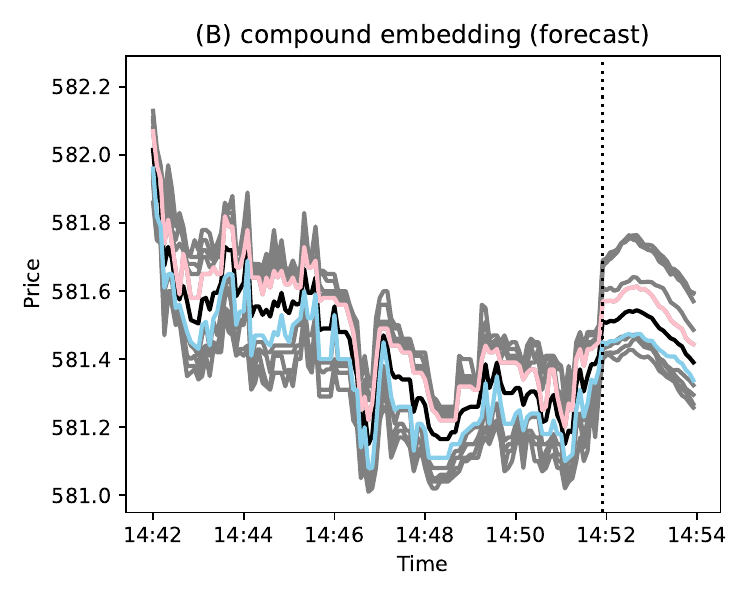} 
    \includegraphics[width=.49\textwidth]{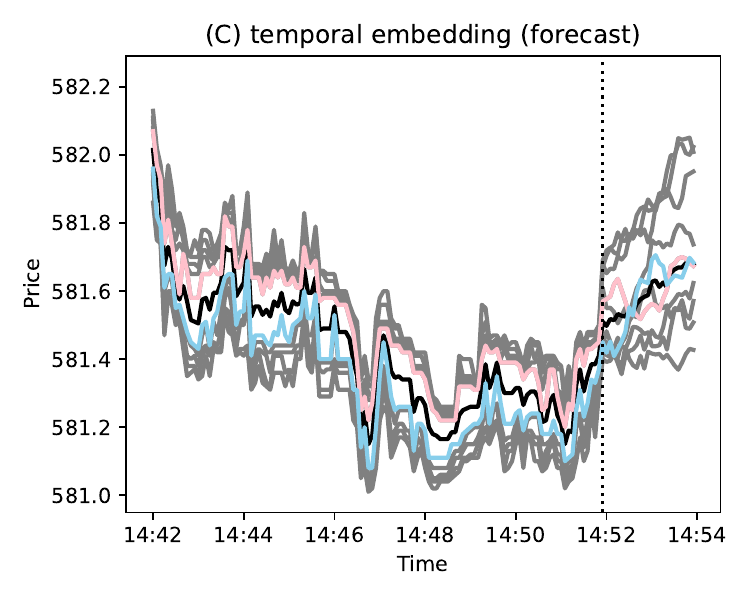} 
    \includegraphics[width=.49\textwidth]{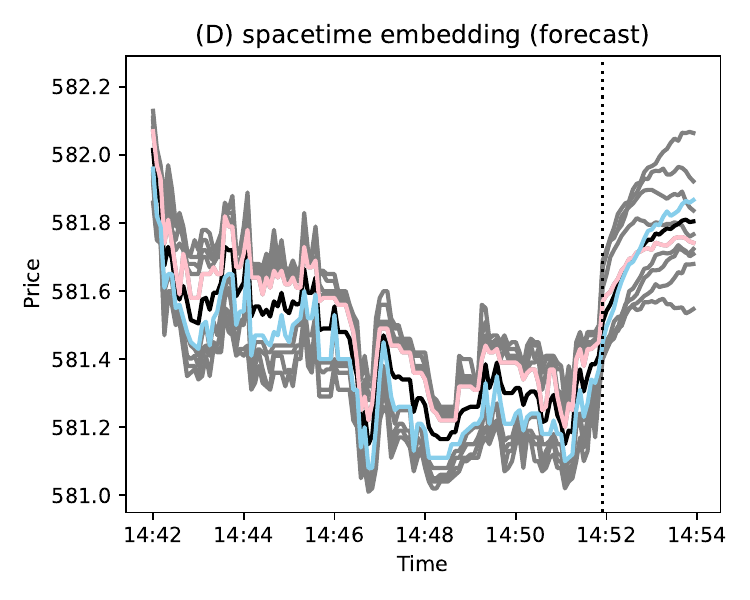} 
    \includegraphics[width=.49\textwidth]{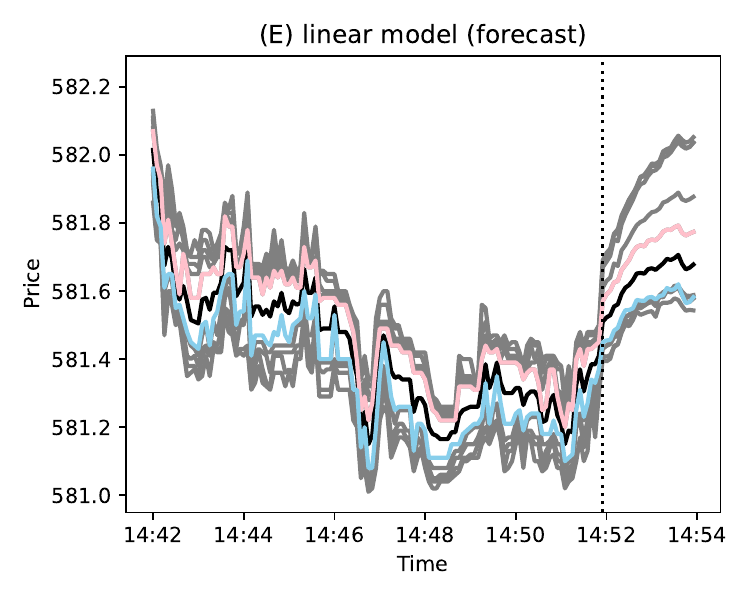} 
    \includegraphics[width=.49\textwidth]{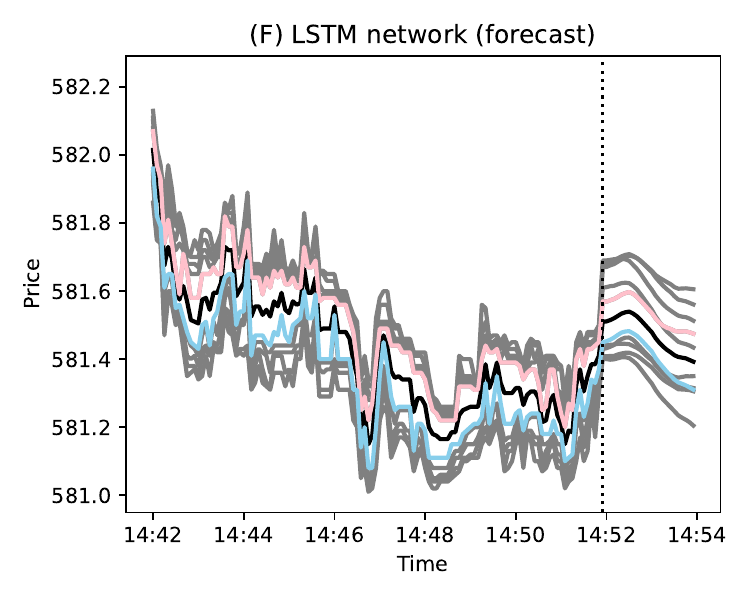} 
    \caption{Price forecasting of a limit order book for AAPL stock: The graphs predict level-5 prices 2 minutes ahead, based on the preceding 10 minutes of context, with the prediction period starting on the right side of the dotted line. Panel A (top left) shows the actual price movements, while panels B, C, and D display the predicted price movements using five forecasting methods: compound multivariate embedding (B), temporal embedding (C), spatiotemporal embedding (D), linear model (E), and LSTM network (F). Multi-level prices, excluding the best prices, are represented in sequential grey lines.}
    \label{fig:AAPL_example}
\end{figure}

\begin{figure} 
    \centering
     \includegraphics[width=.45\textwidth]{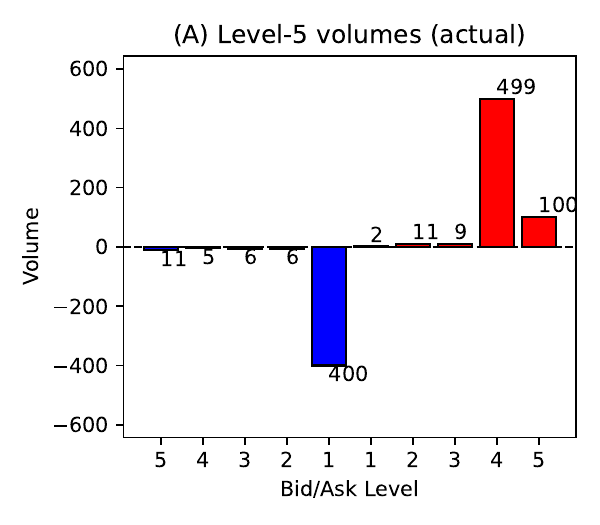} 
    \includegraphics[width=.45\textwidth]{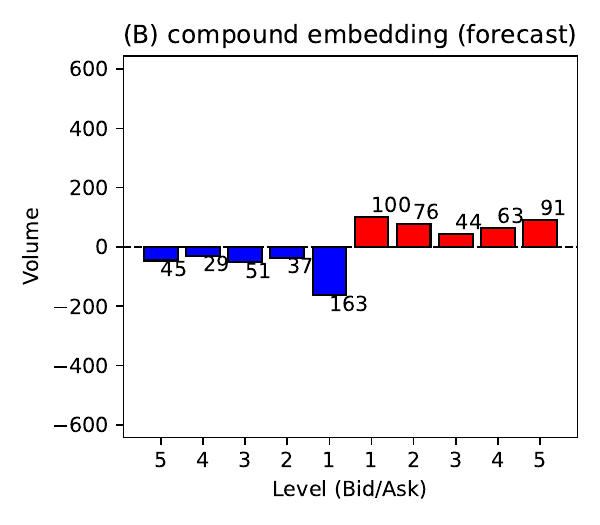} 
    \includegraphics[width=.45\textwidth]{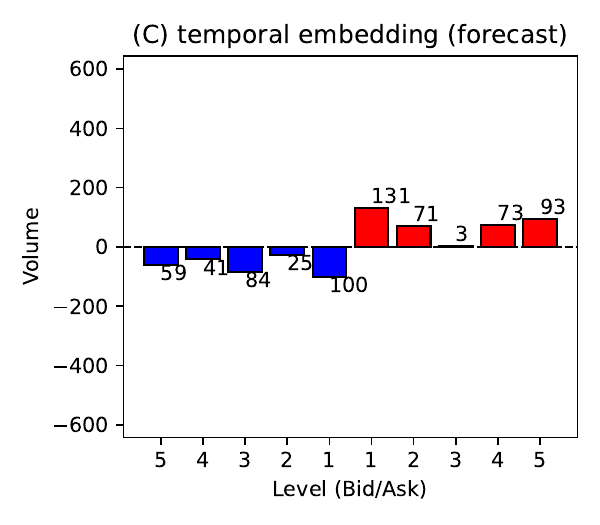} 
    \includegraphics[width=.45\textwidth]{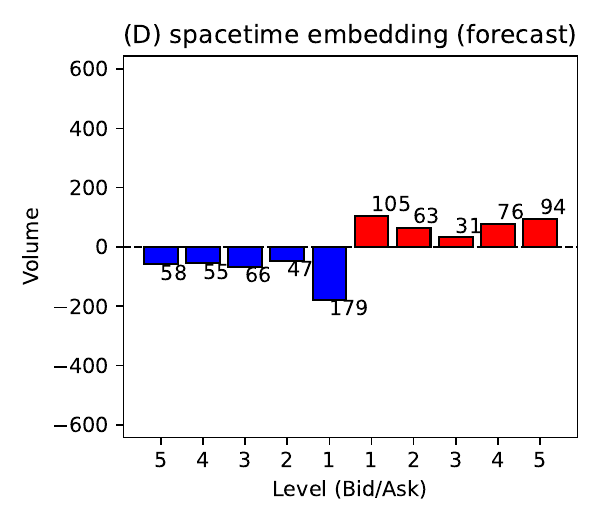} 
    \includegraphics[width=.45\textwidth]{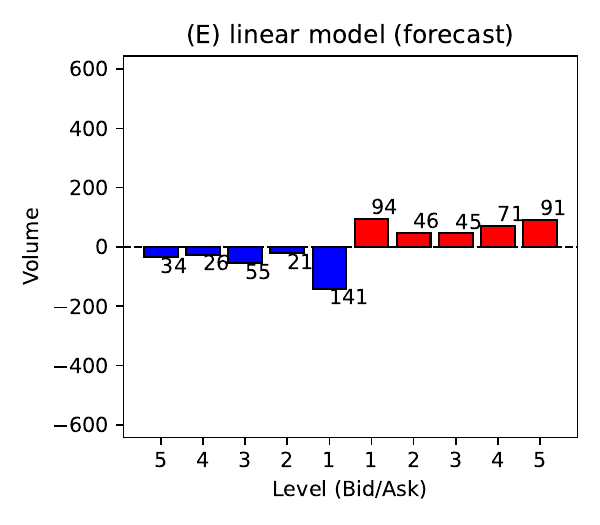} 
    \includegraphics[width=.45\textwidth]{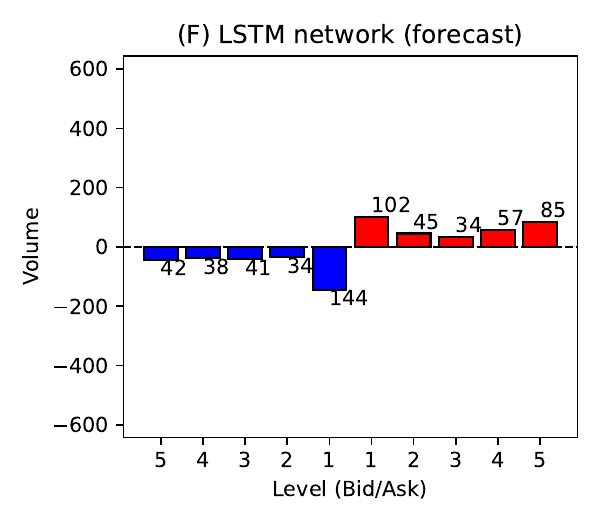} 
    \caption{Volume forecasting of a limit order book snapshot for GOOG stock: The graphs forecast level-5 volumes for the first snapshot of the testing period based on the preceding 10 minutes of context. Panel A shows the actual order sizes, while panels B, C, and D display the predicted volume snapshots using five forecasting methods: compound multivariate embedding (B), temporal embedding (C), spatiotemporal embedding (D), linear model (E), and LSTM network (F).}
    \label{fig:GOOG_example_vol}
\end{figure}

\begin{figure} 
    \centering
     \includegraphics[width=.45\textwidth]{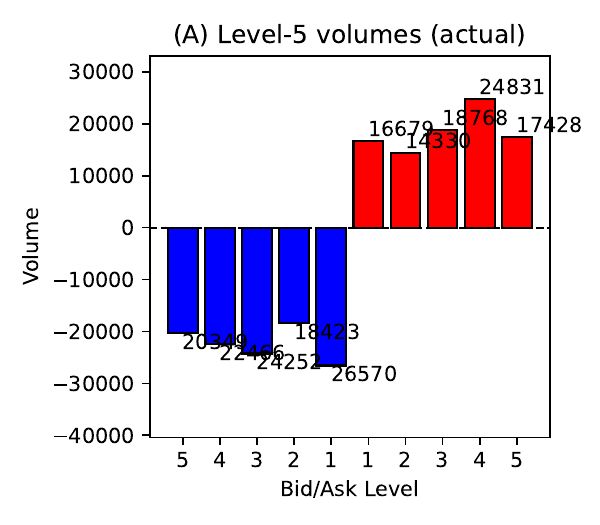} 
    \includegraphics[width=.45\textwidth]{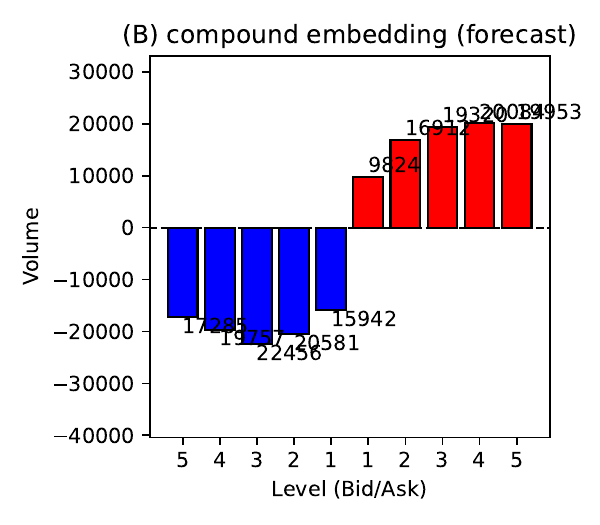} 
    \includegraphics[width=.45\textwidth]{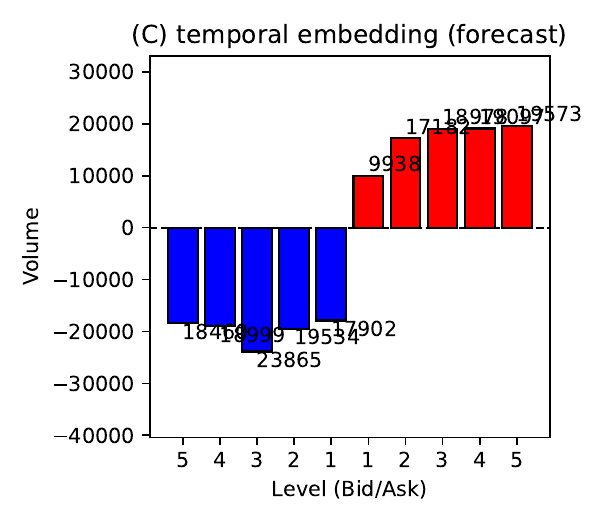} 
    \includegraphics[width=.45\textwidth]{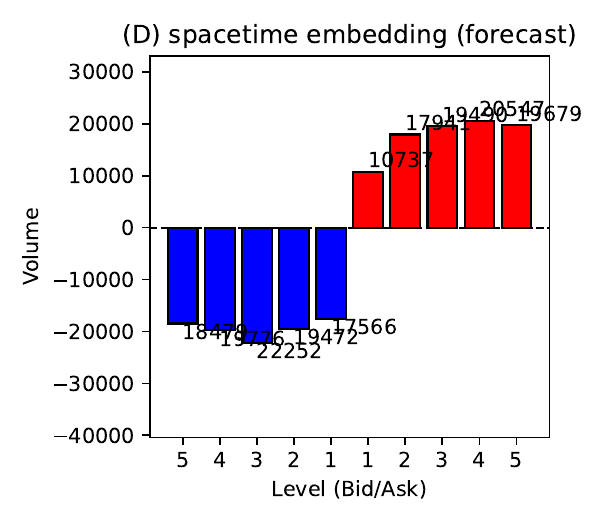} 
     \includegraphics[width=.45\textwidth]{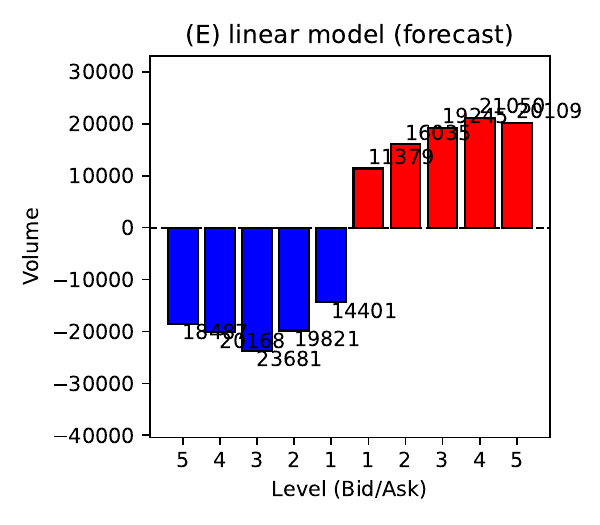} 
    \includegraphics[width=.45\textwidth]{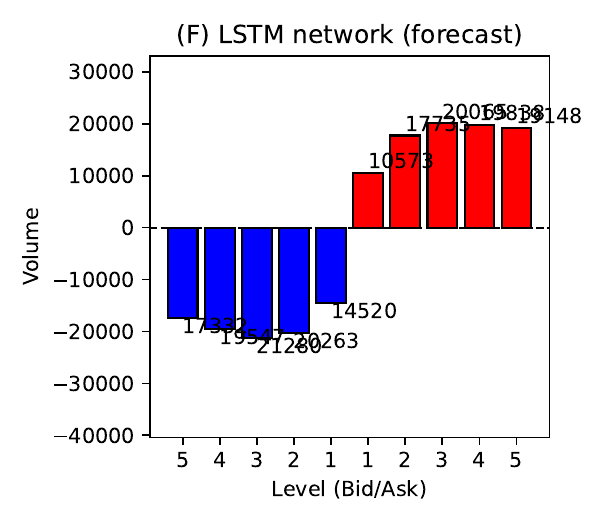} 
    \caption{Volume forecasting of a limit order book snapshot for MSFT stock: The graphs forecast level-5 volumes for the first snapshot of the testing period based on the preceding 10 minutes of context. Panel A shows the actual order sizes, while panels B, C, and D display the predicted volume snapshots using five forecasting methods: compound multivariate embedding (B), temporal embedding (C), spatiotemporal embedding (D), linear model (E), and LSTM network (F).}
    \label{fig:MSFT_example_vol}
\end{figure}

\begin{figure} 
    \centering
     \includegraphics[width=.45\textwidth]{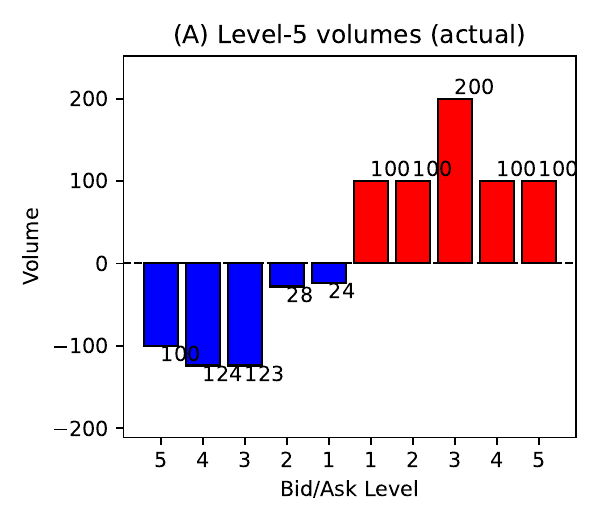} 
    \includegraphics[width=.45\textwidth]{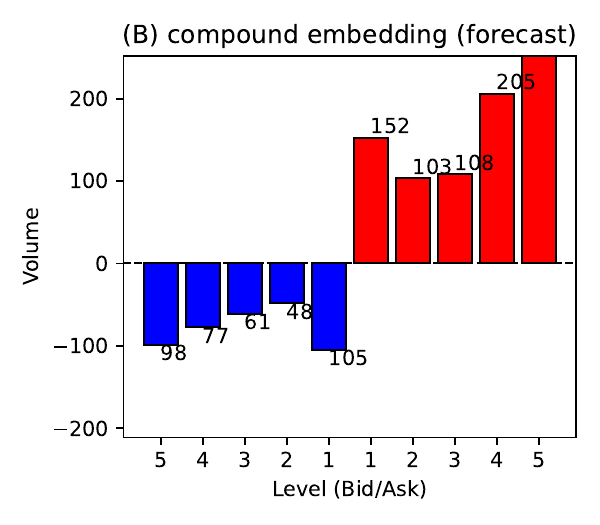} 
    \includegraphics[width=.45\textwidth]{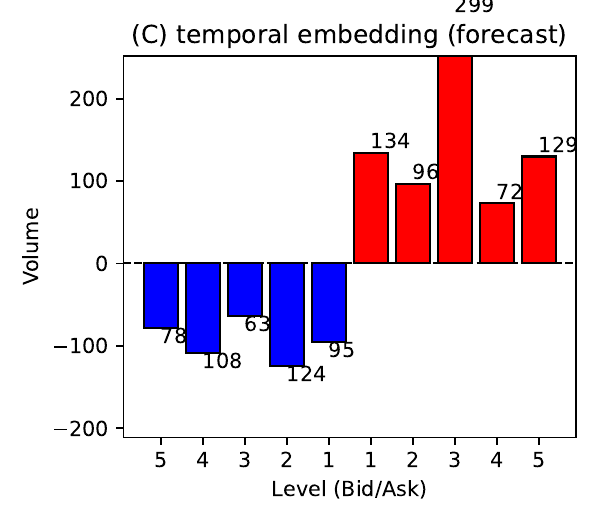} 
    \includegraphics[width=.45\textwidth]{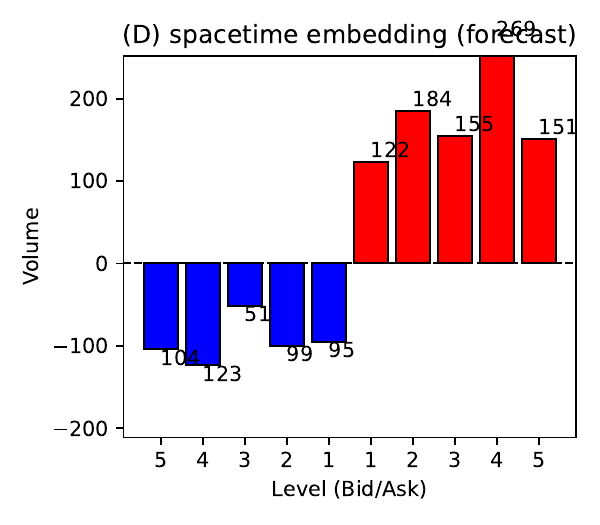} 
     \includegraphics[width=.45\textwidth]{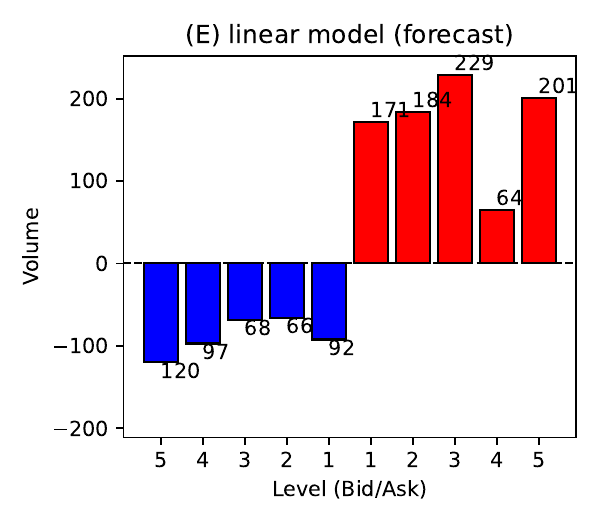} 
    \includegraphics[width=.45\textwidth]{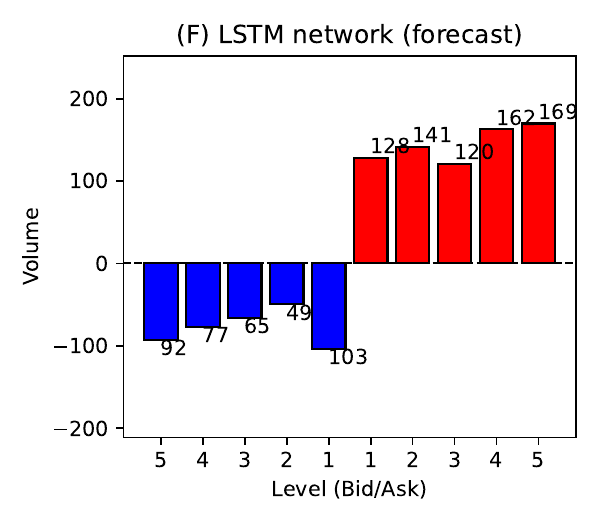} 
    \caption{Volume forecasting of a limit order book snapshot for AAPL stock: The graphs forecast level-5 volumes for the first snapshot of the testing period based on the preceding 10 minutes of context. Panel A shows the actual order sizes, while panels B, C, and D display the predicted volume snapshots using five forecasting methods: compound multivariate embedding (B), temporal embedding (C), spatiotemporal embedding (D), linear model (E), and LSTM network (F).}
    \label{fig:AAPL_example_vol}
\end{figure}

Similarly, in Figure~\ref{fig:AMZN_example_vol}, the volume forecasting results indicate that the compound multivariate embedding method (B) produces a snapshot of volume distributions that align closely with the actual order sizes (A). Differences among the attention-based models are less pronounced for volume predictions than for price forecasts. Additional examples of outputs from all models, including both time series and attention-based approaches, are provided for other stocks, with results shown in Figures~\ref{fig:GOOG_example}, \ref{fig:MSFT_example}, \ref{fig:AAPL_example}, \ref{fig:GOOG_example_vol}, \ref{fig:MSFT_example_vol}, and \ref{fig:AAPL_example_vol},.

These empirical results underscore the effectiveness of the compound multivariate embedding method in capturing the intricate dependencies within the LOB data, leading to more accurate and reliable forecasts for both price and volume.

\section{Discussions}

% Implication 
Our approach to forecasting the entire LOB is fully data-driven, allowing for predictions of both multi-level prices and their respective order sizes at each level. This multi-level perspective expands our analytical horizon beyond mid-price forecasting, where practical applications are often limited. By capturing the granular structure of price levels, our model better reflects real market dynamics, providing a more actionable foundation for order execution.

In addition to price information, market depth—represented by the order sizes across levels—plays a significant role in influencing the LOB’s rapid fluctuations. Events such as the addition, cancellation, or execution of large orders lead to priority shifts and reordering within levels, causing the LOB to dynamically adjust as orders enter or exit different price tiers. Our comprehensive, data-driven approach enables a more accurate depiction of market behavior by integrating the full scope of LOB information. Due to computational resource limitations, our experiments were conducted on the top five levels of the LOB. However, our method is extendable, allowing for deeper depth levels and broader time horizons.

% Limitation
While our model effectively captures dependencies by focusing on multi-level price and volume data alone, one limitation lies in its inability to identify the precise causes of these fluctuations—whether due to new orders, cancellations, executions, or modifications. Using an attention-based architecture, the model adapts in real time to shifts in market context, accurately tracking evolving dependencies within the LOB. However, future work could enhance this by incorporating the specific types of order events, allowing the model to differentiate between sources of LOB changes and better interpret their unique effects on LOB dynamics.

% Potential application in Optimal Order Execution
In practice, the ability to forecast multi-level LOB changes can significantly enhance strategies for optimal execution of large orders by simulating the LOB’s response based on current data. This approach allows traders to anticipate how their trades might affect market depth, spread, and price dynamics, enabling them to adjust order placement to reduce slippage and minimize market impact. By leveraging multi-level insights, they can optimize trade size and timing to prevent adverse price moves and execute in a way that preserves liquidity.

\section{Conclusions}

In this study, we addressed the challenge of forecasting multi-level LOB data, a task that often exceeds the capabilities of conventional time-series forecasting models. Our primary contribution is the development and implementation of a compound multivariate embedding method within advanced sequence-to-sequence models. This method effectively captures the complex interdependencies among various LOB attributes, including order types, features, and levels.
The empirical results demonstrate that our approach consistently outperforms existing multivariate forecasting methods, achieving the lowest forecasting errors while maintaining the ordinal structure of the LOB. These findings highlight the effectiveness of our method for financial time-series forecasting, particularly in the context of high-frequency trading.
As a direction for future work, our method could be extended to address optimal execution problems, with further optimization for real-time forecasting scenarios to enhance its practical applicability in dynamic trading environments.

\pagebreak

\bibliographystyle{ws-ijtaf}
\bibliography{ws-ijtaf}

\newpage
% \appendix 

\FloatBarrier
\beginsupplement 
\section{Supplementary Material} 
% \begin{figure}
%     \centering
%     % \includegraphics[width=\linewidth]
%     \caption{Caption}
%     \label{fig:enter-label}
% \end{figure}

\end{document}